\documentclass[fleqn,usenatbib]{mnras}

\usepackage{newtxtext,newtxmath}

\usepackage[T1]{fontenc}

\DeclareRobustCommand{\VAN}[3]{#2}
\let\VANthebibliography\thebibliography
\def\thebibliography{\DeclareRobustCommand{\VAN}[3]{##3}\VANthebibliography}

\usepackage{graphicx}	
\usepackage{amsmath}	
\usepackage{amssymb}
\usepackage[version=4]{mhchem}
\usepackage{longtable}
\usepackage{threeparttable}
\usepackage{booktabs}
\usepackage{caption}
\usepackage{placeins}
\usepackage{adjustbox}
\usepackage{ragged2e}
\usepackage{xcolor}

\title[]{The physical and chemical structure of Titan's upper atmosphere: energy balance, photochemistry and Jeans escape}

\author[S. Boro Saikia et al.]{
Sudeshna Boro Saikia,$^{1}$\thanks{E-mail: sudeshna.boro.saikia@univie.ac.at}
Jonathan Tennyson,$^{2}$ 
Shuchen Ji,$^{2}$  
Nils-Martin Robeling,$^{1}$
Ivan Stankovi\'c,$^{1}$
\newauthor
Gwena\"elle Van Looveren,$^{1}$
Simon Schleich, $^{1}$
Colin P. Johnstone,$^{3}$
Kristina Kislyakova,$^{1}$
Manuel G\"udel,$^{1}$
\\
$^{1}$Department of Astrophysics, University of Vienna,
              T\"urkenschanzstrasse 17, A-1180 Vienna, Austria\\
$^{2}$Department of Physics and Astronomy, University College London, London, WC1E 6BT, UK\\
$^{3}$ International Institute for Applied Systems Analysis, Schlo\ss{}platz 1, A-2361, Laxenburg, Austria
}

\date{Accepted XXX. Received YYY; in original form ZZZ}

\pubyear{\the\year{}}

\begin{document}
\label{firstpage}
\pagerange{\pageref{firstpage}--\pageref{lastpage}}
\maketitle

\begin{abstract}
Titan's upper atmospheric layers provide a unique opportunity to study how photochemistry, photoionisation, haze microphysics and escape processes shape its atmospheric chemical structure, providing a unique benchmark for understanding these processes in exoplanets. Using a 1D first-principles approach, we extend the Kompot model to simulate Titan's upper atmosphere and investigate its thermal and chemical structure, and Jeans escape. We performed self-consistent simulations of  Titan's upper atmosphere, above the dense haze layer, over a 1D radial grid by solving the coupled hydrodynamics, thermal and photochemical kinetic, and energy balance equations. We considered heating by stellar X-ray and ultraviolet (XUV) and infrared (IR) radiation, and cooling due to radiative emission of methane (\ce{CH4}) and hydrogen cyanide (\ce{HCN}). We also included additional nineteen photochemical and 117 thermal reactions to Kompot's standard chemical network.
  Our modelled atmospheric temperature structures lie within the observed variability of Titan's atmospheric temperature profile. Our simulations reproduce observed nitrogen (\ce{N2}) and \ce{CH4} densities in the upper atmosphere, and abundance of minor chemical species also broadly agree with observationally derived abundances. Near the exobase, our \ce{CH4} mixing ratio overlap with Cassini-Huygens results, and our model \ce{H2} mixing ratios also exhibit broad agreement with observations. Our simulations suggest no significant thermal loss of \ce{CH4}, even under the conditions of an early Sun's XUV environment.
\end{abstract}




\section{Introduction}
Titan, Saturn's largest moon, provides a unique laboratory to study how nitrogen (\ce{N2}) dominated atmospheres originate and evolve under conditions vastly different from Earth's. Furthermore, the co-existence of methane (\ce{CH4}) with \ce{N2} makes it a compelling natural laboratory to study atmospheric disequilibrium processes, such as photochemistry, photoionisation, and photochemically produced aerosols known as hazes. The cascade of chemical reactions associated with these disequilibrium processes leads to the production of complex organic compounds in Titan's atmosphere.  Outside the Solar System, evidence of such atmospheric disequilibrium \citep{tsai_photochemically_2023,beatty_sulfur_2024,jaziri_unraveling_2025} and detection of atmospheric \ce{CH4} \citep{guilluy_gaps_2022,taylor_are_2025, madhusudhan_carbon-bearing_2023} have been only reported in hydrogen-rich massive exoplanets. No clear evidence of an atmosphere rich in \ce{N2} or \ce{CH4} has been detected around smaller exoplanets \citep[e.g.,][]{zieba_no_2023,xue_jwst_2025}. However, the wide diversity of exoplanets observed by the {James Webb Space Telescope (JWST)}, and upcoming next-generation instruments make \ce{CH4} and other hydrocarbon discovery in a widely diverse population of exoplanets a near-future reality. Robust interpretation of such discoveries would require an understanding of how various physical and chemical processes shape the atmospheric chemical and thermal profiles and observable signatures of Titan, the best example in the Solar System of an 
\ce{N2}- and \ce{CH4}-rich atmosphere in disequilibrium. 

While the exact origin of Titan's present-day atmospheric composition is debated, formation of \ce{N2} via photolysis of ammonia (\ce{NH3}) and outgassing of \ce{CH4} are considered to be the most likely mechanisms \citep{tobie_episodic_2006,horst_titans_2017}. Observations taken by the {Cassini-Huygens} mission and ground-based instruments such as {ALMA} have revealed its rich atmospheric chemistry, which includes molecular hydrogen, noble gases, hydrocarbons, cyanides, and oxygen-bearing species \citep[and references therein]{nixon_composition_2024}. Many of these chemical species form precursors to Titan's organic haze, which plays an important role in Titan's atmospheric dynamics and radiative processes \citep{tomasko_rain_2005}. 

Central to Titan's diverse atmospheric chemistry are photodissociation and photoionisation driven by the solar high-energy radiation at X-ray and ultraviolet (XUV) wavelengths. Ionisation in Titan's atmosphere is also driven by solar energetic particles, cosmic rays and particles from Saturn's magnetosphere \citep{cravens_model-data_2009,gronoff_ionization_2009}. These processes have been studied extensively with photochemical kinetics models. Before {Cassini-Huygens} the models were primarily focussed towards neutral chemistry involving \ce{CH4} and \ce{N2} \citep{strobel_photochemistry_1974,yung_photochemistry_1984}. With the discovery of heavier ions in Titan's upper atmosphere, photochemical models are now expanded to include both ion and neutral chemistry \citep{horst_origin_2008,krasnopolsky_chemical_2014,dobrijevic_1d-coupled_2016,vuitton_simulating_2019}. Some models also explicitly couple photochemical kinetics with haze particle formation, linking gas-phase chemistry with aerosol microphysics \citep{lavvas_coupling_2008,lavvas_coupling_2008-1,lavvas_detached_2009}.  
The chemical abundances derived in these models are generally in good agreement with observations. However, existing models cannot yet explain the observations of complex heavier ions and the high \ce{CH4} and \ce{H2} abundance in Titan's upper thermosphere. These discrepancies are closely tied to limitations in our knowledge of heavier ion chemistry, missing reaction rates, and model sensitivity to the assumed atmospheric thermal profile.
 
The temperature structure of an atmosphere plays a critical role in shaping its chemical profile by controlling reaction rates, condensation, and escape of atmospheric gases to space. {\it In-situ} measurements taken by  {Cassini-Huygens' HASI} instrument, as it landed on Titan's surface, revealed a complex atmospheric temperature structure with multiple temperature inversions \citep{fulchignoni_situ_2005}. These measurements are at least a few tens of Kelvin hotter than temperatures derived from {\it in-situ} measurements taken by the Cassini-Huygens' INMS instrument during multiple flybys \citep{snowden_thermal_2013} and remote sensing data \citep{lellouch_intense_2019,thelen_variability_2022}. The observed disagreement between different instruments can be attributed to instrumental sensitivity, seasonal and diurnal variations, and Titan's location within Saturn's magnetosphere \citep{snowden_thermal_2013,nixon_atmosphere_2025}. A range of models have simulated Titan's vertical temperature profile where heating due to solar extreme ultraviolet (EUV) radiation and radiative cooling by molecules are the key mechanisms regulating the temperature structure \citep{yelle_non-lte_1991,muller-wodarg_thermosphere_2000,bell_developing_2014,snowden_thermal_2014}. The exact molecules included in the radiative cooling calculations, however, vary from model to model. Effects of aerosol heating and particle precipitation from Saturn's magnetosphere are also considered in some models as additional energy sources \citep{lavvas_detached_2009,snowden_thermal_2014}. 

However, a self-consistent atmospheric model that integrates atmospheric chemistry, temperature structure calculated from first principles, and thermal escape rate has not been applied to study Titan's upper atmosphere. This matters as thermal structure and chemical composition is tightly coupled via radiative heating and cooling, and chemical reaction kinetics, which in turn affects the thermal escape of atmospheric gas and atmospheric evolution. A self-consistent upper atmospheric model could answer some of the outstanding questions on Titan's upper atmospheric temperature structure, chemical composition gradients, atmospheric escape, and long-term evolution. Such a unified self-consistent approach is also critical for exoplanet atmospheres where observations are sparse and limited.

In this work, we update the 1D upper atmospheric code Kompot \citep{johnstone_upper_2018}, originally developed to study the upper atmospheres of Earth and Venus, to simulate the thermo-chemical strcuture of Titan's upper atmosphere. We include the feedback between chemical reactions and energy balance in our self-consistent  model of Titan's upper atmosphere and benchmark our model against remote sensing and {\it in-situ} observations of Titan.  We also calculate the Jeans escape rate within our self-consistent simulations. 
This paper is organised as follows: in Section 2 we provide a general description of the Kompot code, followed by a description of the model updates implemented in this work. The results and discussion are provided in Section 3, and the conclusions are outlined in Section 4.

\section{Model description}
The simulations presented in this work were carried out with an updated version of the 1D thermo-chemical code Kompot \citep{johnstone_upper_2018}, as described in the following sub-sections. The standard Kompot code solves the equations governing hydrodynamics, chemical kinetics, and energy transport across a radial grid to self-consistently calculate the thermal and the chemical profile of planetary upper atmospheres, and the thermal loss of individual chemical species. 
The lower boundary of the simulation can be arbitrarily set at any altitude in the atmosphere, and the upper boundary lies at the exobase. The lower boundary is usually set above any cloud or haze layers as their effects are currently not included in the model. 
The gas mixture in the model is split into ions, neutrals and electrons with their own separate temperatures, and the vertical structure of the atmosphere is calculated as the balance between thermal, chemical, and hydrodynamic processes. 

The Kompot code calculates the atmospheric thermal structure by solving the energy equations, which include heating and cooling terms, thermal conduction, and energy exchange between ions, electrons and neutrals. In addition to direct heating by the stellar XUV and infrared (IR) radiation field, the model accounts for electron heating by elastic collisions with nonthermal electrons, exothermic heating and Joule heating. The cooling is driven by radiative losses by molecules and atoms which include \ce{CO2}, \ce{H2O}, \ce{NO}, \ce{H3+}, and \ce{O}. 
The chemical profile is calculated using a network of chemical reactions, coupled with advection, and eddy and molecular diffusion. While molecular diffusion is calculated self-consistently, eddy diffusion is treated as a free parameter. The original network consists of 503 chemical reactions containing 63 neutral and 30 ion species. This includes 56 photoreactions and 7 photoelectron reactions.

Within the simulation domain of a Kompot model, the atmosphere expands and contracts in response to the changes in the gas temperature and composition, and the code allows for a dynamic calculation of the exobase. The Jeans escape rate for each species is calculated at the exobase following the prescription of \citet{bauer_planetary_2004}. A full description of the Kompot code, including its validation against Earth and Venus, can be found in \citet{johnstone_upper_2018}. For an additional discussion on the Jeans escape rate, we refer the reader to \citet{van_looveren_airy_2024}. Kompot has also been applied to study the hydrodynamic escape of \ce{H2O}  \citep{johnstone_hydrodynamic_2020} and \ce{N2} \citep{johnstone_extreme_2019} atmospheres. More recently, we also carried out a grid of Kompot atmospheric simulations to investigate the atmospheric retention of \ce{N2} and \ce{CO2} rich atmospheres on small rocky exoplanets around a diverse stellar population \citep{van_looveren_airy_2024,van_looveren_habitable_2025}.

In the following sub-sections we describe the model updates carried out in this work, which include extending the energy calculations to include radiative cooling due to \ce{CH4} and \ce{HCN}, and the chemical network to include nitrile, hydrocarbon and ammonia chemistry. 
These updates extend the code's applicability to a broader class of planetary atmospheres, including those found on sub-Neptunes and outer Solar System bodies.

\subsection{Thermal structure}\label{sec:thermal}
Thermal emission of \ce{CH4} and \ce{HCN} plays an important role in determining Titan's atmospheric heat balance \citep{yelle_non-lte_1991,bell_simulating_2010,rezac_rotational_2013}.  For the \ce{CH4} cooling rate, we consider non-local thermodynamic equilibrium (non-LTE) effects by applying the formalism of \citet{yelle_non-lte_1991}, where level populations are determined from statistical equilibrium for a selected set of ro-vibrational bands. In addition, we computed the \ce{CH4} cooling rate using a spectroscopic line list under the LTE assumption, for comparison with the band-averaged  non-LTE formulation used in the \citet{yelle_non-lte_1991} scheme. For \ce{HCN}, however, we only consider cooling rates calculated under the LTE assumption.

The total non-LTE \ce{CH4} cooling rate, assuming a photon escape probability of 1, is given by

\begin{equation}
    {Q_\mathrm{NLTE}(T)=\sum _{i=1}^4 n_{u,i} hc{\tilde{\nu}_i} A_i},
\end{equation}
where $h$ is the Planck constant in erg s, $c$ is the speed of light in cm s$^{-1}$, and $n_{u,i}$, $\tilde{\nu}_i$ and $A_i$ are the population of the upper energy level in cm$^{-3}$, wave number in cm$^{-1}$ and Einstein A-coefficient in s$^{-1}$ respectively for the band $i$ at 7.7, 3.3, 2.3 and 1.7 $\mu$m.  To estimate the level population $n_{u,i}$ we follow the two-level approximation

\begin{equation}
    n_{u,i} = n_{\ce{CH4}} \dfrac{C_{lu,i}}{(C_{ul}+A_i)},
\end{equation}
where $n_{\ce{CH4}}$ is the number density of \ce{CH4}, $C_{ul}$ is the collisional de-excitation rate and $C_{lu,i}$ is the collision excitation rate for each band; $u$ stands for the upper and $l$ stands for the lower energy level. The collisional dexcitation rate is related to the de-excitation probability $P_{ul}$ through 
\begin{equation}
   C_{ul} = n_{\ce{N2}} ~P_{ul} ~\sigma ~\sqrt{\dfrac{8~k_B~T}{\mu}},
\end{equation}
where $n_{\ce{N2}}$ is the number density of \ce{N2} in cm$^{-3}$, $\sigma$ is the effective collision cross section in cm$^{-2}$, $k_B$ is the Boltzmann constant in erg K$^{-1}$,  $T$ is the gas temperature in K, and $\mu$ is the reduced mass in g. The collisional excitation rate is estimated following
\begin{equation}
    C_{lu,i} = C_{ul}~\exp{\left(\dfrac{-h\nu_i}{k_B~T}\right)},
\end{equation}

In this formulation each rotational band is represented by a single effective transition with the same de-excitation probability $P_{ul}=1.4\times10^{-6}$ \citep[see Tables 1 and 2 in][]{yelle_non-lte_1991}. Furthermore, the statistical weights for the upper and lower levels are considered to be identical. Including a more extensive statistical equilibrium calculation for the level population $n_u$ is currently not feasible, as it requires collision rate data for all vibrational-rotational levels with different atmospheric constituents (e.g., \ce{N2}, \ce{H2}, \ce{e-}) which are largely unavailable in current databases. Hence, we adopt the approximation of \citet{yelle_non-lte_1991} to account for non-LTE \ce{CH4} cooling due to collisions with \ce{N2}, the dominant atmospheric gas in Titan, and compare the results against the LTE calculation described below.

To determine the LTE cooling rate of \ce{CH4} and \ce{HCN} we use the approach of \citet{tennyson_radiative_2016} and calculate the cooling rate per molecule, $\Lambda (T)$ in units of erg s$^{-1}$ sr$^{-1}$ molecule$^{-1}$. 
\begin{equation}
\centering
    \Lambda(T) = \dfrac{1}{4\pi Z(T)} \sum_{i,j}A_{ij } ~h~c~\tilde{\nu}_{ij}~g_i ~\exp\left(\dfrac{-c_2 E_i}{T}\right),
    \label{lambda_eq}
\end{equation}
where $Z(T)$ is the partition function, the sum is over all transitions with $i$ and $j$ being the upper and lower states of each transition respectively, $A_{ij}$ is the Einstein A-coefficient (s$^{-1}$), $h=6.626\times 10^{-27}$ erg s is the Planck's constant, $c$ is the speed of light in cm s$^{-1}$, $\tilde{\nu}_{ij}$ is the wave number of the emitted photon in cm$^{-1}$, g$_i$ is the total state degeneracy, $c_2=1.43877736$ cm K is the second radiation constant, $E_i$ is the energy of state $i$ in cm$^{-1}$, and $T$ is the temperature in K \citep[see][for a detailed description]{tennyson_radiative_2016}. 

To calculate $\Lambda(T)$, we use molecular data from the ExoMol database \citep{tennyson_data_2013,yurchenko_exomol_2014}. We take the \ce{CH4} cooling rate directly from the ExoMol database, which is based on the older \lq \lq 10to10\rq\rq ~line list of \citet{yurchenko_exomol_2014}. We recomputed the cooling function using the newer \lq\lq MM\rq\rq ~line list for methane from \cite{yurchenko_exomol_2024}, which showed no
significant changes. Methane does not have a permanent dipole moment and is therefore an inefficient cooler at Titan temperatures. Conversely, \ce{HCN} is a strongly dipolar system; we calculate the cooling rate directly using the \lq\lq Harris\rq\rq\ ExoMol \ce{HCN} line list \citep{barber_exomol_2014}. This and other cooling functions are in the process of being made available via the new database of cooling functions hosted on the ExoMol website.

From $\Lambda (T)$ we calculate the total LTE cooling rate for each species following, 
 \begin{equation}
\centering
    {Q_\mathrm{LTE}(T) = 2 \pi ~n~\Lambda (T)},
    \label{Q_cool}
\end{equation}
where $n$ is the number density of a given species, and the factor 2$\pi$ accounts for integration over a hemisphere, assuming radiation is emitted in the upward direction only. Since we do not include absorption and re-emission of radiation in our model, we assume this simplification provides a reasonable estimate of the net LTE cooling rate. 

\subsection{Chemical network}
We updated the chemical network in Kompot with nineteen photoreactions (Appendix \ref{chem_appendix}, Table~\ref{table:photoreaction}), and 117 thermal reactions involving only neutrals (Table~\ref{table:chem_neutral}). The wavelength-dependent cross-sections were taken from \textsc{PHIDRATES} \citep{huebner_photoionization_2015}. Due to the non-availability of photodissociation cross sections of \ce{C4H2}, \ce{C4H4} and \ce{C2N2}, we used the UV absorption cross sections (all measured below 2400 \AA) from the \textsc{MPI-Mainz UV/VIS Spectral Atlas} \citep{keller-rudek_mpi-mainz_2013} instead as a proxy and assumed a quantum yield of unity. We consider such an approximation to be reasonable, as photodissociation is expected to dominate over other effects at these wavelengths. Furthermore, any uncertainties introduced by this assumption are expected to have a negligible impact on the model results as this approximation is only applied to the three minor species in Titan's \ce{N2}- and \ce{CH4}-rich atmosphere. 

These newly added chemical reactions are based on the \citet{yung_photochemistry_1984} network, with additional reactions from \citet{hebrard_how_2009}. We also implemented the reactions involving ammonia ($\ce{NH3}$) from the full $\ce{S-N-C-H-O}$ network of the open-source photochemical kinetics code VULCAN \citep{tsai_comparative_2021}. This includes all reactions with $\ce{NH2}$, $\ce{N2H}$, and $\ce{N2H2}$, as these are important intermediates in $\ce{NH3}$ chemistry. The temperature range and rate coefficients of all newly added reactions were adopted from the KInetics Database for Astrochemistry \citep[KIDA,][]{wakelam_kinetic_2012}. For reactions with multiple rate constants in the literature, we followed KIDA guidelines and selected their recommended values. If no such recommendation was available, we used rate constants from cited works in KIDA, with preference given to values used in models benchmarked against Titan \citep[e.g.,][]{hebrard_how_2009}.
We calculated the energies associated with each thermal reaction as the difference between the total formation enthalpies of the products and those of the reactants, using enthalpy values from the NIST Chemistry Webbook \citep{linstrom_nist_2001} and the Argonne Active Thermochemical Tables \citep{ruscic_active_2022}. We used the formation enthalpy ($\Delta_\mathrm{f}H^\circ$) of each chemical species at the standard temperature of 298.15\ K.
Furthermore, for some thermal reactions, we adopted updated product channels to reflect our current understanding of Titan's atmospheric composition. As an example, the channel involving the reactants $\ce{OH}$ and $\ce{CH3}$, which leads to the production of $\ce{CO}$ in \citet{yung_photochemistry_1984}, is not included here. Experiments by \citet{de_avillez_pereira_temperature_1997} and \citet{baulch_evaluated_2005} suggest alternative channels, none of which produce $\ce{CO}$. We include the channels proposed by \citet{wong_evolution_2002} and the channels from \citet{baulch_evaluated_2005} in our chemical network. New calculations are also available for reactions involving a number of other chemical species \citep{pearce_crahcn-o_2020}. However, incorporating these rates would require a broader assessment of the available reaction rates throughout the network to ensure a consistent treatment across all reactions. We therefore retain the reaction rates selected following the methodology described above and defer a systematic evaluation of new reactions and their associated reaction rates to future work.

\subsection{Model setup and input parameters}\label{setup}
  We perform four model simulations to explore the sensitivity of our results to the choice of eddy diffusion parameterisation and to LTE versus non-LTE assumptions in the cooling rate calculations. We benchmark the model results against upper-atmospheric observations of Titan. Finally, we use the observation-constrained model to investigate the temporal evolution of Titan’s thermal and chemical structure, and Jeans escape rate.
 
 In our model setup, we adopt a mass of 0.0225 $M_\oplus$, a radius of 0.404 $R_\oplus$ and an orbital distance of 9.45 au to the Sun for Titan\footnote{\url{https://www.esa.int/Science_Exploration/Space_Science/Cassini-Huygens/Facts_about_Titan}}.
 We set the lower boundary of our simulation domain at an altitude of 450 km, which is just below Titan's mesopause\footnote{Titan's reported mesopause altitude vary with instrument and season. We use the definition used in \citet{vuitton_chapter_2024} and  \citet{nixon_atmosphere_2025}},  thus spanning the entire thermosphere from the mesopause to the exobase. The choice of lower boundary also avoids Titan's dense haze layers, thus minimising the influence of haze microphysics and radiative effects that are currently not included in our model. The zenith angle was set to an arbitrary angle of 20$^\circ$ as the zenith angle has a negligible effect on the simulated atmospheric profiles. A reference homopause, which separates the well-mixed atmosphere from the stratified atmosphere, was initialised at an altitude of 875 km. The temperature and the number density, held constant at the lower boundary of the simulation domain, were set to 180 K and 4.0$\times 10^{14}$ cm$^{-3}$ respectively. These values were adopted to be consistent with observational data \citep{fulchignoni_situ_2005,lellouch_intense_2019,vuitton_simulating_2019,thelen_variability_2022}. During the simulation, however, the atmospheric structure and the homopause are evolved self-consistently until a steady-state is reached.

We set the chemistry at the lower boundary of our simulation with a gas mixture composed of \ce{N2}, \ce{CH4}, \ce{CO}, and \ce{Ar}. At 450 km (lower boundary of our simulation grid), Titan's atmospheric \ce{CH4} abundance is constrained by observational data, but no corresponding measurements are known for \ce{N2}. Hence, we set the \ce{CH4} mixing ratio ($\chi_{\mathrm{CH_4}}$) to 0.02 \citep{shemansky_cassini_2005,maltagliati_titans_2015}. We also initialise a $\chi_\mathrm{CO}$ of 5.0$\times 10^{-5}$ \citep{serigano_isotopic_2016}. Recent Titan observations with JWST show that the \ce{CO} abundance remains constant throughout the mesosphere and lower thermosphere with an abundance of 55$\pm$5 ppm \citep{nixon_atmosphere_2025}. \ce{CO} production in Titan is attributed to the influx of \ce{O} and \ce{OH} from an external source, as described by \citet{horst_origin_2008}. While our chemical scheme includes chemical reactions involving \ce{O}, \ce{OH}, and \ce{CO}, our model currently does not account for any external influx of material at the top. Hence, we set the \ce{CO} abundance at the base of the simulation to the observed values, rather than modelling its production self-consistently. We introduce 50 ppm of \ce{Ar} at the lower boundary of our simulation as a tracer of vertical transport in our atmospheric model, and to compare against Cassini-Huygens measurements. Although the radiogenic\ce{^{40}Ar} is observed in Titan's thermosphere \citep[see][and references therein]{vuitton_chapter_2024}, and is commonly used as a tracer, we use bulk argon in our model. Given that \ce{^{40}Ar} is the most abundant isotope of argon, and it doesn't participate in Titan's \ce{N2}-\ce{CH4} chemistry, we do not introduce isotopic separation. We adjust the \ce{N2} mixing ratio ($\chi_{\ce{N2}}$) to 0.9799 to ensure that the sum of the four mixing ratios is equal to one. The choice of \ce{N2} abundance is in agreement with the $\geq$ 0.95 reported in the literature \citep{vuitton_chapter_2024}.

\begin{figure}
    \centering
    \includegraphics[width=0.95\linewidth]{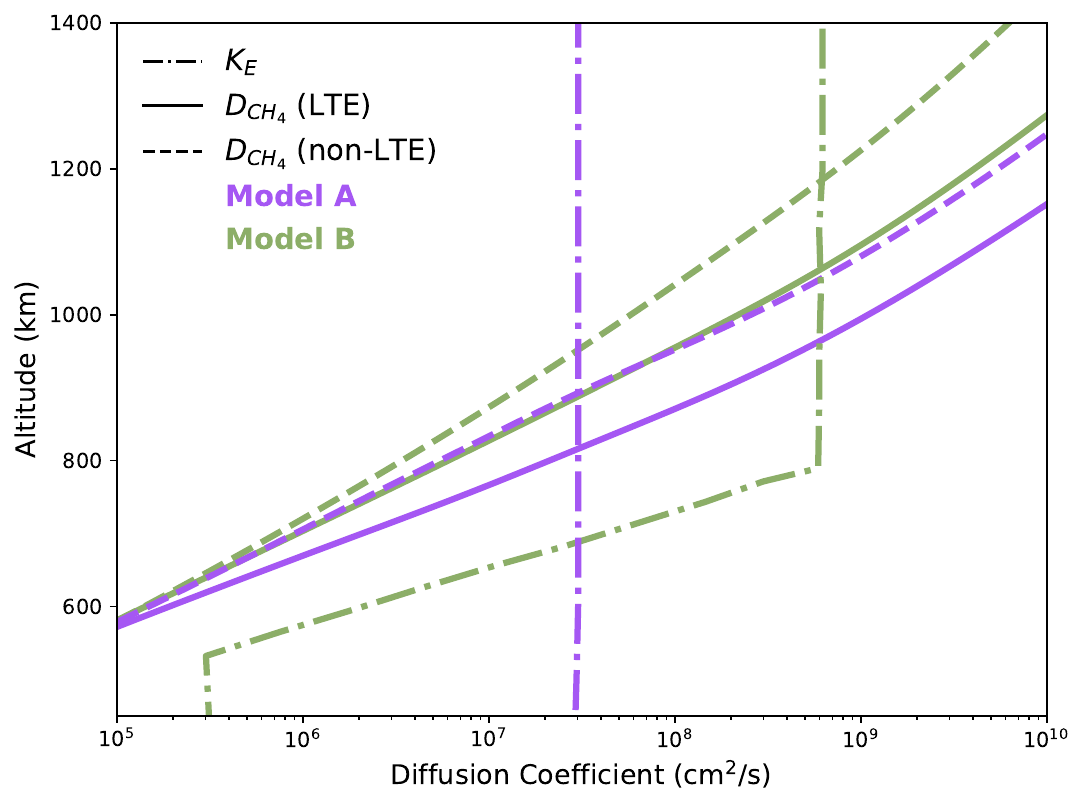}
    \caption{Eddy and molecular diffusion coefficients applied in this work. The dash-dotted lines represent the two $K_E$ profiles used in this work, \citealt{vuitton_simulating_2019} (violet) and \citealt{krasnopolsky_chemical_2014} (green). The solid lines mark the molecular diffusion coefficient of \ce{CH4} calculated for Models A and B, where LTE \ce{CH4} cooling is used in the energy balance equation. The molecular diffusion profiles for the corresponding non-LTE models are shown as dashed lines.}
    \label{fig:diffusion}
\end{figure}

The primary external energy source in our thermal calculation is the present day solar XUV spectrum, which we take from \citet{claire_evolution_2012}. The IR spectrum used here is a simple blackbody with an effective temperature equal to that of the Sun. Joule heating is not expected to have a dominant effect on Titan's thermal profile \citep{snowden_thermal_2014}, and is hence not included in our simulations. As for the radiative cooling sources, we consider \ce{HCN} and \ce{CH4} as they are the dominant emitting molecules in Titan's upper atmosphere. The diffusive flux of particles and conductive flux of energy in our model are controlled by the eddy and the molecular diffusion coefficients. We calculate the molecular diffusion coefficient ($D_{\mathrm{mol}}$) self-consistently, but the eddy diffusion coefficient ($K_E$) is a free parameter.

Titan's eddy diffusion profile is constrained in the literature by finding a best fit model to the chemical abundance profile of nitriles, hydrocarbons and noble gases observed in Titan's atmosphere, with observations ranging from the stratosphere to the thermosphere. Since $K_E$ is not a well-constrained parameter, we tested our simulations against two different eddy diffusion profiles (Models A and B), as shown in Fig.~\ref{fig:diffusion}. The eddy profile adopted in Model A, while constant in the pressure ranges considered in this work, is based on the asymptotic function fitted by \citet{vuitton_simulating_2019} to {Cassini-Huygens} data. The model of \citet{vuitton_simulating_2019} is one of the most comprehensive photochemical models post Cassi-Huygens, and its adopted eddy diffusion profile is broadly consistent with previous studies \citep{moreno_abundance_2012,loison_neutral_2015}. We therefore adopt this $K_E$ profile as our baseline case. The eddy diffusion profile of Model B was taken from \citet{krasnopolsky_chemical_2014}, which uses a three point profile, with $K_E$ on average an order of a magnitude larger than the profile used in Model A, as shown in Fig.~\ref{fig:diffusion}. The divergence of Model B from Model A can be attributed to the choice in the number of parameters and observational data points used to constrain $K_E$, with Model B representing an extreme end of the currently available eddy diffusion estimates for Titan's upper atmosphere. Together, Models A and B are intended to capture the range of uncertainty in the currently available estimates of eddy diffusion coefficients in Titan's upper atmosphere. 

Figure~\ref{fig:diffusion} also shows the  molecular diffusion profiles of \ce{CH4} ($D_{\ce{CH4}}$) calculated within each model.   The $D_{\ce{CH4}} \mathrm{(cm^2s^{-1})}$ was calculated using the relation, 
\begin{equation}
    D_{\ce{CH4}} = \dfrac{\alpha_{\ce{CH4}} \times 10^{17} T^{S_{\ce{CH4}}}}{N},
\end{equation}
where $T$ and $N$ represent the temperature and the total number density of the gas. The two coefficients, $\alpha_{\ce{CH4}}$ and $S_{\ce{CH4}}$, were taken from tabulated values for a \ce{N2}-rich atmosphere in \citet{banks_aeronomy_1973}. As shown in Fig.~\ref{fig:diffusion}, the molecular diffusion profile intersects with the eddy diffusion profile at a different altitude in each model. This intersection marks the homopause, which is at an altitude of 800 to 1050 km for the two LTE models. The homopause for Model A (non-LTE) lies at around 900 km, where as the homopause for Model B (non-LTE) can be found at approx. 1200 km. The precise location of the homopause in a steady-state Kompot simulation is sensitive to the eddy diffusion coefficient, as it affects the density profile, and thus the molecular diffusion coefficient.

\section{Results and Discussion}\label{sec:results}
\begin{figure*}
    \centering
    \includegraphics[width=0.9\linewidth]{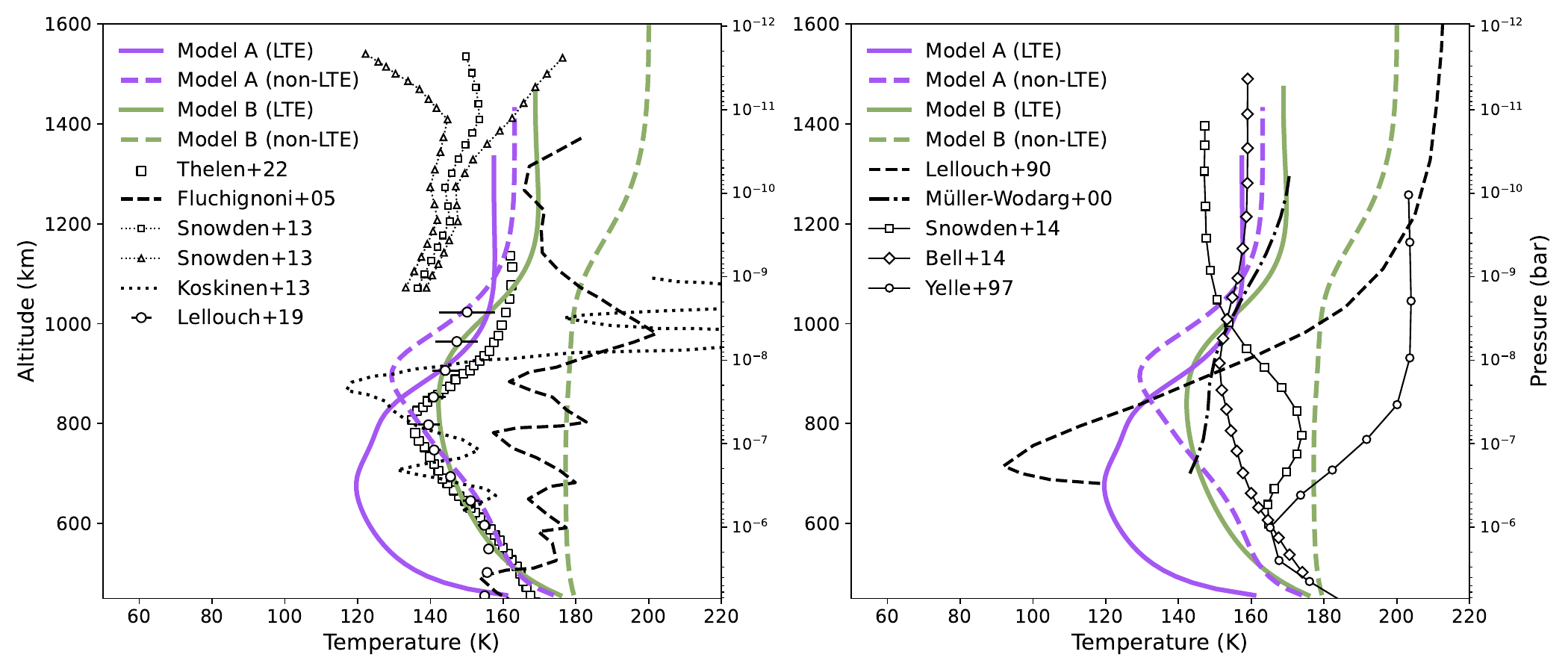}
    \caption{Thermal structure of our four atmospheric models (same colour scheme as in Fig.~\ref{fig:diffusion}). The temperature in K is shown on the x-axis, the altitude in km is shown on the left y-axis, and the pressure in bar is shown on the right y-axis. {\it Left:} Temperatures derived from {ALMA} are marked by squares and circles (\citealt{thelen_variability_2022} and \citealt{lellouch_intense_2019}), from HASI are shown by the dashed line (\citealt{fulchignoni_situ_2005}), and from {UVIS} are shown by the dotted line (\citealt{koskinen_mesosphere_2011}). Globally averaged temperatures from {INMS} data are shown by the square markers with dotted lines \citep{snowden_thermal_2013}. The uncertainties associated with the temperature profiles used to calculate the average profile of \citet{snowden_thermal_2013} are shown by the triangles with dashed lines. {\it Right:} Temperature profiles from thermospheric models in the literature are shown by the black lines. The solar minimum model of \citet{lellouch_intense_2019} is shown by the dashed line, the \citet{yelle_yelle_1997} Titan engineering model is marked by the circles with the solid line, the \citet{muller-wodarg_thermosphere_2000} model is shown by the dashed and dotted line, the \citet{snowden_thermal_2014} model is shown by the square markers with the solid line, and the \citet{bell_developing_2014} model is shown by the diamonds with the solid line.}
    \label{fig:PTprofile}
\end{figure*}

\subsection{Thermal structure}\label{results_cool}
The simulated temperature profiles of our four models are shown in Fig.~\ref{fig:PTprofile}, where the left panel shows a comparison against observational data and the right panel shows a comparison against several existing upper atmospheric models. {While the two LTE models exhibit a similar shape for the thermal profile}, the absolute temperature values depend heavily on the choice of $K_E$ in the model. The temperature structure of the non-LTE Model A is similar to the LTE model B. When compared against the LTE Model A, however, the non-LTE Model A is around 40 K hotter at the lower thermosphere, around 20 K cooler at the middle thermosphere near the homopause, and a few K hotter at the upper thermosphere. The non-LTE Model B is consistently hotter than the other three models throughout the thermosphere. It also exhibits an almost isothermal behaviour, with a 20 K increase in temperature at the upper thermosphere near the exobase.

In the left panel of Fig.~\ref{fig:PTprofile}, we compare our model results against the temperature profiles derived from data taken by the HASI \citep{fulchignoni_situ_2005}, the UVIS \citep{koskinen_mesosphere_2011}, and the INMS \citep{snowden_thermal_2013} instruments aboard Cassini-Huygens, and ALMA \citep{lellouch_intense_2019,thelen_variability_2022}. Titan's observationally derived temperature profiles show discrepancy across instruments due to instrumental sensitivity, spatial coverage, and seasonal and diurnal variations in Titan \citep{nixon_atmosphere_2025}. The HASI temperature profile is inferred from {\it in-situ} density measurements and shows the hottest temperatures throughout the thermosphere. \citet{koskinen_mesosphere_2011} estimated the temperature  from \ce{CH4} number densities derived by UVIS during two Cassini flybys (T41 and T53). The temperatures from T41 are in good agreement with the measurements of the HASI instrument (not shown here to avoid overcrowding the figure). However, the temperature derived from the T53 fly-by exhibits at least 20 K higher temperature at mid-to-upper thermosphere but much lower temperatures at the mid-to-low thermosphere. Using \ce{N2} densities measured by the INMS instrument, \citet{snowden_thermal_2013} also estimated Titan's thermospheric temperature structure, which is at least 20 K cooler than other known measurements. The average profile from 32 Cassini passes (square markers), along with the upper and lower bounds (triangular markers) on the uncertainties associated with the derived temperature profiles of \citet{snowden_thermal_2013} are shown in Fig.~\ref{fig:PTprofile}. The temperature profiles retrieved from ALMA \ce{HCN} and \ce{CO} observations span multiple epochs, and the best-fit temperature profiles from \citet{lellouch_intense_2019} and \citet[2015 epoch]{thelen_observations_2024} are shown here, which fall in between the  temperature profiles derived using the different Cassini-Huygens instruments.   

Depending on the model, our results show varying levels of agreement with the observationally derived temperature profiles. At the mid-to-lower thermosphere (below 800 km), Model A (LTE) is cooler than all observations. However, Model B exhibits strong agreement with the ALMA and the UVIS data. The discrepancy between the LTE Model A simulations and the observations could be due to the missing aerosol heating in our models, which can increase the temperature by approximately 20 K near the bottom of our simulation domain \citep{lavvas_detached_2009}. However, as shown by Model A (non-LTE), assumption of non-LTE \ce{CH4} cooling increases the temperature at the lower thermosphere, exhibiting excellent agreement with ALMA observations. At similar altitudes, Model B (non-LTE) is hotter than all other models, but in a better agreement with HASI measurements. At the mid-to-upper thermosphere (above 800 km), the simulated temperature of all models lie within the observed variability. While all four models exhibit some level of agreement with one or multiple sets of observations, Model A (non-LTE) and LTE Model B show a better agreement with the observations, in particular with ALMA. 

For completeness, we also compare our results against existing upper atmospheric models on the {right} panel of Fig.~\ref{fig:PTprofile}. 
The key differences between our models and other works are the treatment of the stellar XUV heating and the radiative cooling. While Kompot performs a full self-consistent calculation of XUV heating, all other models in the literature apply a simplified approach where the stellar radiation-driven heating is approximated by a parametric heating efficiency factor. The efficiency factor is a free parameter that determines the fraction of the incoming stellar radiation that is converted to heat in the atmosphere. Additionally, the stellar radiation in the X-rays is ignored in the literature models, and only the EUV radiation is considered. The models also differ in terms of the radiative cooling treatment, where they don't always include the same set of molecules in the radiative loss calculation. 

The literature temperature profiles that deviate the most from our results are, Titan's engineering model developed by \citet{yelle_yelle_1997} and the model described in \citet{lellouch_titans_1990}.  We show the average temperature profile from \citet{yelle_yelle_1997} and the solar minimum case from \citet{lellouch_titans_1990} in the right panel of Fig.~\ref{fig:PTprofile}. Solar EUV heating is incorporated via a heating efficiency factor in both studies. In the 1D atmospheric model of \citet{lellouch_titans_1990}, the heating efficiency is parametrised by assuming a linear relationship between the efficiency factor and wavelength. The \citet{yelle_yelle_1997} model is an empirical model based on the thermospheric model of \citet{yelle_non-lte_1991} where the EUV heating is treated in a similar manner as \citet{lellouch_titans_1990}. These two models also differ in the treatment of radiative cooling. The \citet{lellouch_titans_1990} model incorporates radiative cooling due to \ce{CH4} and \ce{C2H2}, where the calculation is carried out per molecule with a dilution factor to account for non-local thermodynamic equilibrium (non-LTE) effects in the upper atmosphere.  The \citet{yelle_non-lte_1991} calculations include non-LTE radiative cooling of \ce{CH4}, \ce{C2H2} and \ce{C2H6} under the two-level approximation, and LTE radiative cooling due to \ce{HCN}. In addition to the treatment of XUV heating and radiative cooling, differences in the input boundary conditions, since these two models were developed before the launch of Cassini-Huygens, could also explain some of the observed discrepancy between our results and these two models in the literature.

The thermospheric temperatures in our suite of models are in a relatively good agreement with the models of \citet{muller-wodarg_thermosphere_2000}, \citet{snowden_thermal_2014} and \citet{bell_developing_2014}. However, the temperature structures of these three models deviate from our results at the mid-to-lower thermosphere.  This deviation could be primarily attributed to the differences in the complexity of the physical processes included in the various models.
All three models treat the stellar EUV heating via an efficiency factor. The simulations of \citet{muller-wodarg_thermosphere_2000} account for radiative cooling by \ce{HCN}, but unlike our model, they do not include any \ce{CH4} cooling. The \citet{snowden_thermal_2014} model, which is based on the \citet{yelle_non-lte_1991} model, includes aerosol heating and an additional efficiency term to account for magnetic particle heating. Considering Titan's location within Saturn's magnetosphere, non-thermal heating is considered to be another key heating process in Titan's atmosphere, a process that we don't account for in our model. The thermal profile of \citet{snowden_thermal_2014} is much hotter (up to 30 K) at mid-to-low thermosphere, which could be attributed to the inclusion of aerosol heating in their model, which is not included in our simulations. Our model B (non-LTE) is a few K hotter than the \citet{snowden_thermal_2014} model, but enhanced vertical transport rather than the cooling or the heating rate is responsible for these temperatures. This is supported by the lower temperature in Model A (non-LTE) which employs the same non-LTE cooling as Model B (non-LTE), but differs in terms of the adopted $K_E$ profile.  
 The \citet{bell_developing_2014} model includes \ce{HCN} cooling and a fixed magnetic heating term in their thermal calculations. The authors also simulated three different scenarios by varying the diffusion coefficient and an imposed \ce{CH4} escape rate. Out of model simulations described in \citet{bell_developing_2014}, the model with the lowest imposed \ce{CH4} escape rate and higher homopause altitude most closely approximates the observed mixing ratios of \ce{CH4} and \ce{HCN}. Hence, we only show this model in Figure~\ref{fig:PTprofile}, which qualitatively agrees with our temperature structure.

As shown in Fig.~\ref{fig:PTprofile}, simulating the thermal structure throughout Titan's upper atmosphere using numerical models is a challenging task. This difficulty is compounded by variability between data taken by different instruments at different phases of Titan's seasonal and diurnal cycles. The data were also collected during the different phases of the solar activity cycle, resulting in variations of the received stellar XUV flux. Our self-consistent treatment of the heating and the cooling terms yields thermospheric temperatures that are well within the range of observed variability, and captures the average temperature structure of Titan's themosphere. The comparable agreement achieved by models incorporating either enhanced eddy diffusion (Model B) or non-LTE radiative cooling (non-LTE Model A) highlights a degeneracy between these mechanisms when considering the thermal structure alone.
A detailed discussion on the different heating and cooling terms that shape the temperature structure in our models can be found in Appendix~\ref{sec:heatcool}.

\subsection{Chemical composition}

\subsubsection{Dominant species: \ce{N2} and \ce{CH4} }
\begin{figure}
    \centering
    \includegraphics[width=0.5\textwidth]{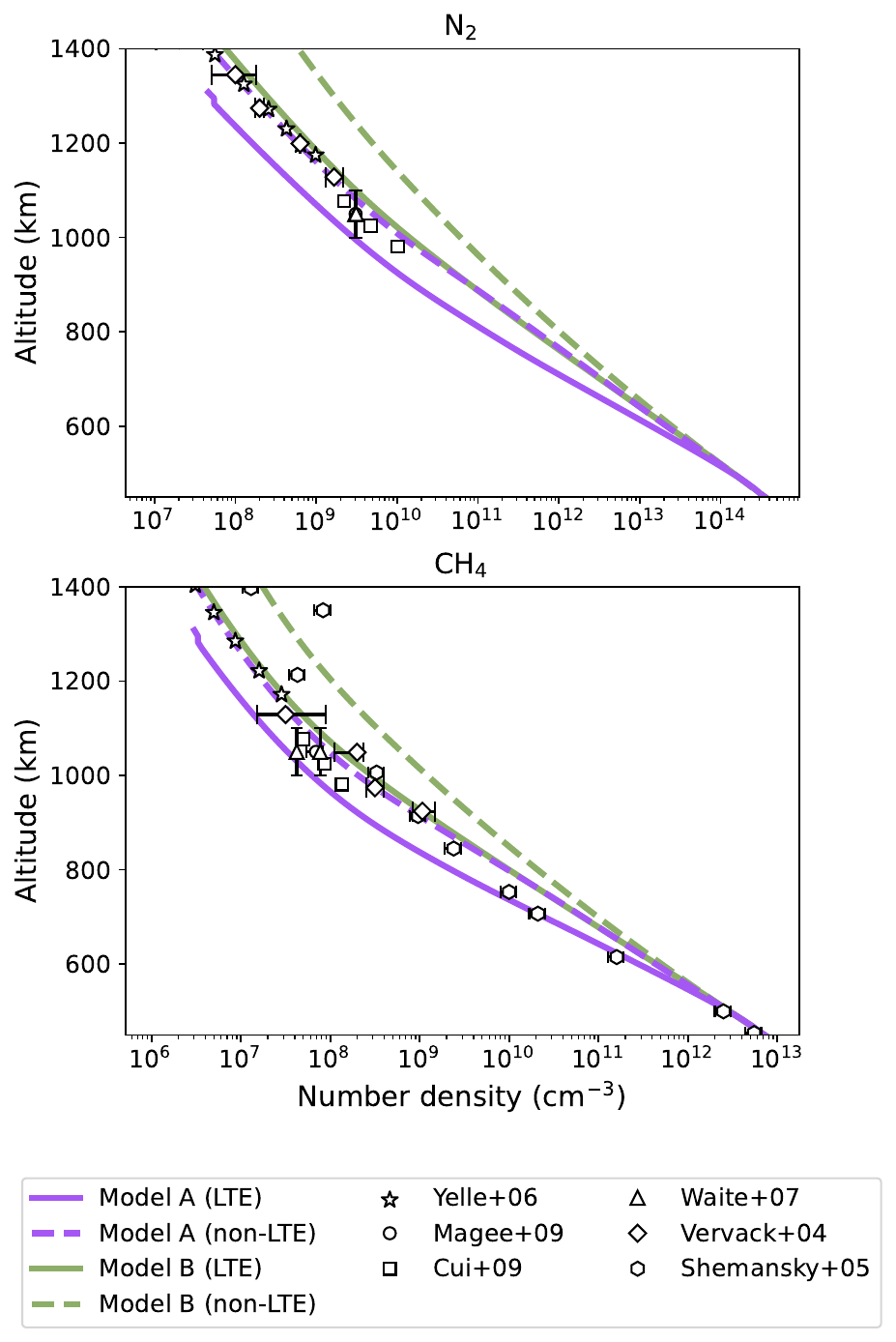}
    \caption{Number densities (cm$^{-3}$) of \ce{N2} and \ce{CH4} as a function of altitude (km). The colours remain the same as in Fig.~\ref{fig:diffusion}. The markers represent data taken by Cassini-Huygens and Voyager I.}
\label{fig:nodensity}
\end{figure}

The simulated number densities of the two most abundant atmospheric species, \ce{N2} and \ce{CH4}, are compared against observations in Fig.~\ref{fig:nodensity}. The number density profiles vary within an order of magnitude between the two LTE models, with both \ce{N2} and \ce{CH4} densities increasing with the eddy diffusion coefficient in our models. The non-LTE Model A overlaps the LTE models, but Model B (non-LTE) shows consistently higher densities at the middle and the upper thermosphere. This increase in density from the non-LTE Model A to B correlates with the increase in $K_E$.

As shown in Fig.~\ref{fig:nodensity},  Titan's \ce{N2} density measurements are confined to the upper thermosphere \citep[also see Fig.~2 in][]{vuitton_chapter_2024}, and are derived from data taken by Voyager I and Cassini-Huygens' INMS instrument. The number densities shown here were obtained from three separate analyses of INMS data \citep{cui_analysis_2009,magee_inms-derived_2009, waite_new_2007}. To maintain clarity, only the tabulated data from \citet{magee_inms-derived_2009} are shown. These measurements are marked by the data points at 1000 km and they are in strong agreement with the two LTE models and Model A (non-LTE). INMS data were also analysed by \citet{yelle_vertical_2006}, where the authors derived the number densities from both the inward and outward trajectories of Cassini-Huygens. Only the outbound leg is shown here for comparison against our model, which is in excellent agreement with Model A (non-LTE) and the LTE Model B, and within an order of magnitude with  LTE Model A. Additionally, we compare our results against the \ce{N2} values derived by \citet{vervack_new_2004} from Voyager I observations, where our model results show the same trend as with the \citet{yelle_vertical_2006} results.

The observationally derived \ce{CH4} number densities are taken from Voyager I, Cassini-Huygens' INMS and UVIS instruments, leading to a full data coverage of our simulation domain, as shown in Fig.~\ref{fig:nodensity}. Closer to 1000 km, our model results (two LTE models and the non-LTE Model A) fall well within the variability and any uncertainty associated with the INMS measurements analysed by \citet{magee_inms-derived_2009,cui_analysis_2009} and  \citep{waite_new_2007}. The densities derived by \citet{yelle_vertical_2006} from INMS data, however, are slightly higher at the upper thermosphere which are in excellent agreement with Model A (non-LTE) and Model B, and within an order of magnitude with LTE Model A. The \ce{CH4} densities derived from Voyager I \citep{vervack_new_2004} fall at the middle of our cluster of four models. The \ce{CH4} densities derived from the UVIS data analysis of \citet{shemansky_cassini_2005} are in strong agreement with our LTE models throughout the thermosphere, except at the very top near the exobase. When comparing the non-LTE models, Model A shows a better agreement with this dataset at the lower thermosphere, but Model B is in a better agreement with \citet{shemansky_cassini_2005} at the upper thermosphere. Given that the \ce{CH4} number densities derived by \citet{shemansky_cassini_2005} at these altitudes also deviate from the densities derived from INMS data, it could be attributed to the UVIS's poor sensitivity at the upper thermosphere \citep{koskinen_mesosphere_2011}. New analysis of UVIS data by \citet{yelle_titan_2021} does not show these very high densities near the exobase. Hence, we consider our model results, except Model B (non-LTE), to be in good agreement with observed \ce{CH4} densities throughout the thermosphere, with Model A (non-LTE) and LTE Model B exhibiting the strongest agreement with the data.

\subsubsection{Minor species}
\begin{figure}
    \centering
     \includegraphics[width=0.445\textwidth]{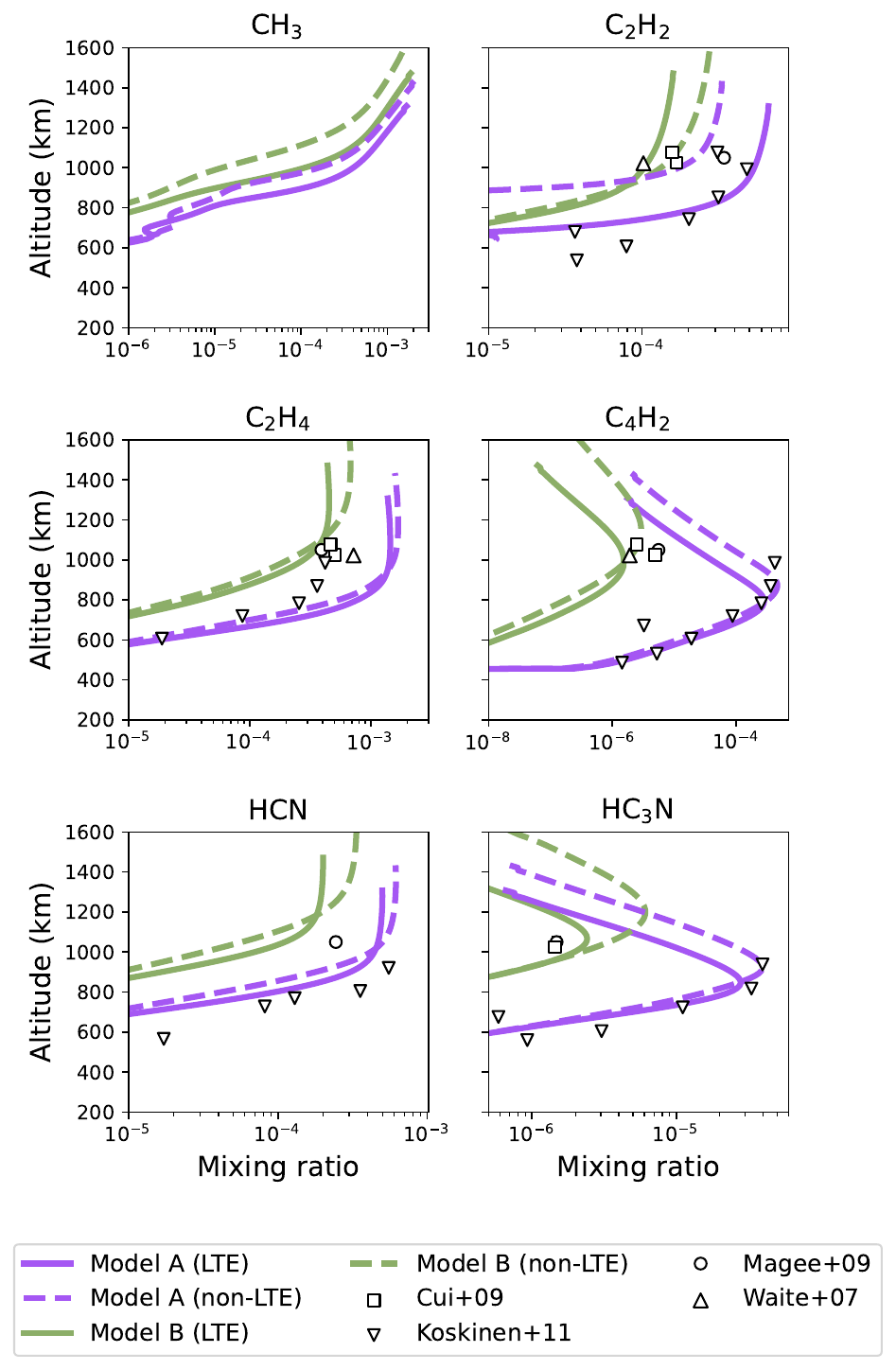}
\caption{Mixing ratio vs altitude of six minor species, \ce{CH3}, \ce{C2H2}, \ce{C2H4}, \ce{C4H2}, \ce{HCN} and \ce{HC3N}. Our six models have the same colour scheme as in Fig.~\ref{fig:diffusion}. The open symbols represent the mixing ratios obtained from various observations.}
\label{fig:minor_species}
\end{figure}

\begin{figure}
    \centering
    \includegraphics[width=0.5\textwidth]{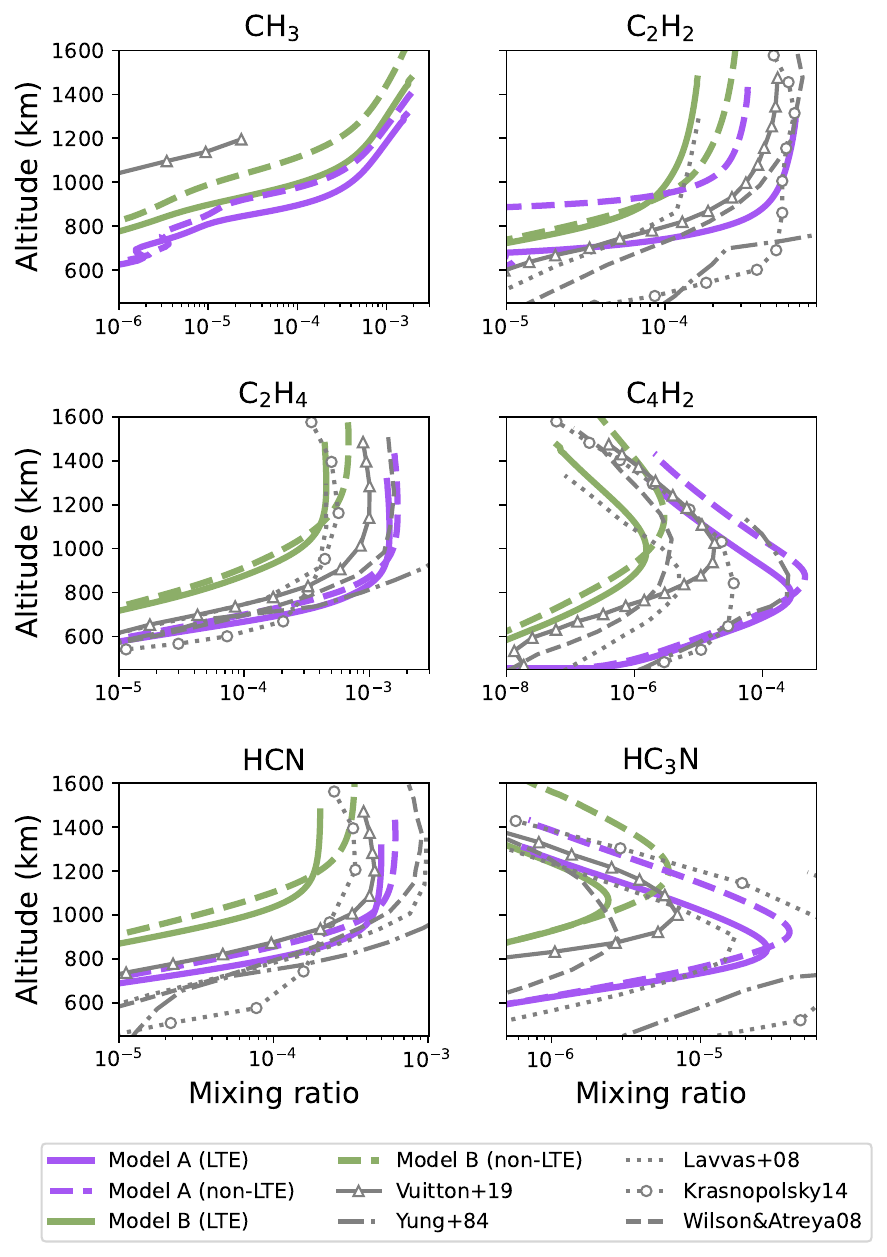}
    \caption{Same as Fig.~\ref{fig:minor_species} but model results from the literature are compared instead of observational data. The dashed line represents the model results of \citet{wilson_current_2004}, the dashed and dotted lines mark the model of \citet{yung_photochemistry_1984}, the dotted lines are from \citet{lavvas_coupling_2008-1}, the dotted lines with open circles show the results from \citet{krasnopolsky_chemical_2014}, and the solid line with triangles mark the \citet{vuitton_simulating_2019} model. }
    \label{fig:photochem_compare}
\end{figure}

Figures~\ref{fig:minor_species} and \ref{fig:photochem_compare} show the mixing ratios of a limited selection of six minor species present in Titan's atmosphere, \ce{CH3}, \ce{C2H2}, \ce{C2H4}, \ce{C4H2}, \ce{HCN}, \ce{HC3N}. We selected these species because they were identified in Titan's atmosphere by multiple Cassini-Huygens visits. The methyl radical (\ce{CH3}) is also included as it was recently discovered in Titan's atmosphere using JWST's MIRI instrument \citep{nixon_atmosphere_2025}, although no mixing ratio has been reported so far.  The observational data shown here were obtained from Cassini-Huygens' INMS \citep{magee_inms-derived_2009,cui_analysis_2009,waite_cassini_2004} and UVIS \citep{koskinen_mesosphere_2011} instruments.

In our models, the mixing ratio profiles of these minor species are primarily sensitive to the choice of eddy diffusion. Depending on the species, the choice of \ce{CH4} cooling function has a minor impact. The only exception is \ce{CH3} whose abundance profile does not show any strong correlation with the choice of $K_E$ or \ce{CH4} cooling function. For the other five species with Cassini-Huygens data, agreement between instruments also vary from species to species. Amongst the four models, the LTE model A exhibits a similar trend as the \ce{C2H2} abundance derived from UVIS data. The two non-LTE models and the LTE Model B are in a better agreement with the INMS derived \ce{C2H2} abundances. The observed \ce{C2H4} and \ce{HCN} abundances show the weakest variability between the instruments. The observed \ce{C2H4} lies in between our choice of models. However, the \ce{HCN} abundance is in a better agreement with Models A (LTE and non-LTE). For \ce{C4H2} and \ce{HC3N}, the LTE and the non-LTE model B align with the INMS data but these two models have a significantly lower abundance than the UVIS data. Model A (LTE and non-LTE) are in strong agreement with the UVIS dataset. These two models also loosely agree with the INMS data with a slightly higher abundance. 
 When the observed variability in measurements is considered, our ensemble of models capture the average trend in the abundance profile of these minor species.

A comparison of our model results against the photochemical models of \citet{yung_photochemistry_1984}, \citet{wilson_current_2004}, \citet{lavvas_coupling_2008-1}, \citet{krasnopolsky_chemical_2014} and \citet{vuitton_simulating_2019} is shown in Fig.~\ref{fig:photochem_compare}.   
The \citet{yung_photochemistry_1984} model was developed pre-Cassini-Huygens, so the model assumptions are not as well constrained as the more recent models discussed here. As an example, the chemical kinetics data and reaction pathways have been updated with new laboratory measurements in the last few decades. \citet{wilson_current_2004} use the thermal profile from \citet[as shown in the right panel of Fig.~\ref{fig:PTprofile}]{yelle_yelle_1997} as an input to calculate the chemical structure. They test their model against a wide range of $K_E$, and their nominal model is shown here. In addition to photodissociation due to stellar EUV, the model also accounts for cosmic ray deposition. The model of \citet{krasnopolsky_chemical_2014} comprises both neutral and ion chemistry, cosmic ray influx, water delivery at the top of the atmosphere, and magnetospheric electrons and ions. The \citet{lavvas_coupling_2008-1} results shown here are based on their photochemical/microphysics model that accounts for aerosol formation and its impact on gas chemistry. Finally, \citet{vuitton_simulating_2019} simulates the full atmosphere, extending from the ground to the exobase and includes neutral as well as ion chemistry in their model.

The \ce{C2H4}, \ce{C4H2}, and \ce{HCN} abundance in the literature models show a similar structure as our ensemble of models. This suggests that the key reaction pathways connected to these species are captured by all current photochemical models, including ours. However, the profiles of \ce{C2H2} and \ce{HC3N} show varying levels of overlap between our results and other models. While the \ce{C2H2} and \ce{HC3N} profiles from our models broadly agree with the results of \citet{vuitton_simulating_2019}, the abundance of \ce{CH3} is higher in our models. Since our chemical network is not as extensive as some of the networks in the literature, in a future study we will investigate the  impact of extended neutral and ion chemistry on the minor species abundance in Titan's upper atmosphere.

\subsubsection{Argon abundance}
\begin{figure}
    \centering
    \includegraphics[width=0.5\textwidth]{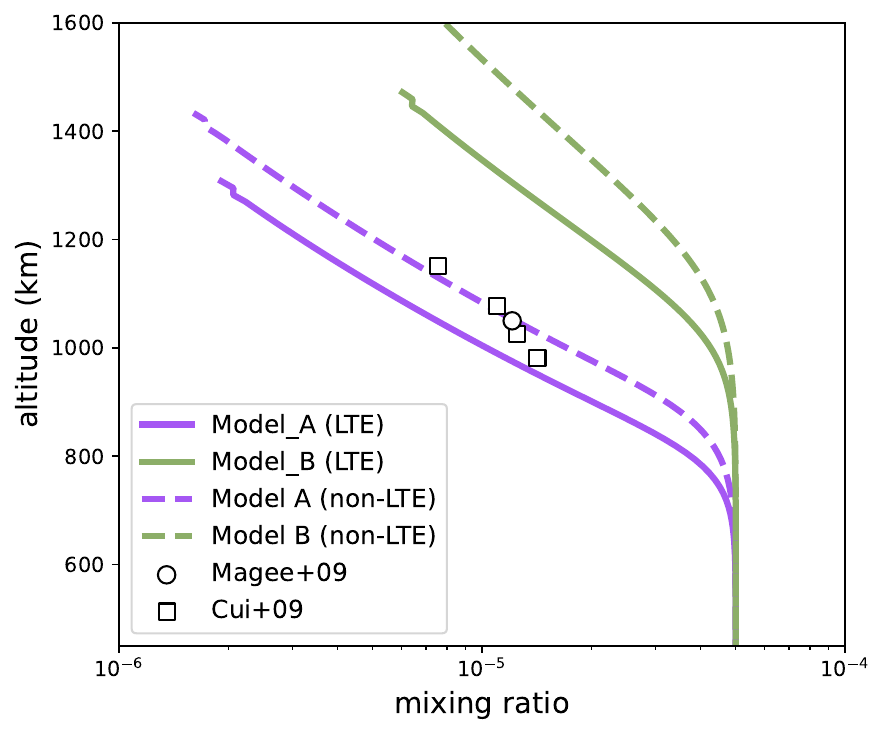}
    \caption{Argon mixing ratio. Same colour scheme  as Fig.~\ref{fig:diffusion}. The markers represent observational data from Cassini-Huygens.}
    \label{fig:argon}
\end{figure}
The hydrocarbon and nitrile abundances discussed in the previous section exhibit limited discriminatory power between the models, as different models agree with different subsets of the observations. In contrast, \ce{Ar}, being chemically inert, is more directly influenced by the atmospheric thermal structure and transport processes. Figure~\ref{fig:argon} shows the mixing ratio profile of argon which strongly correlates with the choice of eddy diffusion profile. In general, the argon abundance decreases with an increase in altitude. However, at the upper layers of the thermosphere the abundance is higher for models with stronger vertical mixing. For the same $K_E$ profile, the non-LTE models exhibit higher argon abundance when compared to the LTE models, but the effect is not as strong as the choice of $K_E$. The Ar abundance measured by the INMS instrument show good agreement with our weaker $K_E$ models, with the closest match provided by Model A (non-LTE).

\subsection{Atmospheric escape}
 Table~\ref{tab:escape_rates} lists the top five chemical species by Jeans escape rate in our four models. In Model A (LTE), \ce{H}, \ce{H2}, \ce{H+}, \ce{H2+} and \ce{H3+} exhibit the largest escape rates where as in Model B (LTE), \ce{CH4} and \ce{CH3} have stronger Jeans escape than \ce{H2+} and \ce{H3+}. In Model A (non-LTE) the species list changes to \ce{H}, \ce{H2}, \ce{H+}, \ce{H2+}, and \ce{CH4}, whereas in the case of Model B (non-LTE), \ce{N} rather than \ce{H2+} appears in the list of five species with the strongest Jeans escape.
 
Titan's inferred \ce{H} escape rate from Cassini-Huygens observations by \citet{hedelt_titans_2010} is 2.8$\times 10^3$  g~s$^{-1}$. However, \citet{muller-wodarg_titans_2014} estimate an escape rate of 1.7$\times 10^3$ g~s$^{-1}$, highlighting the difficulty in constraining Titan's \ce{H} escape rate. The \ce{H2} escape rate is however, calculated over multiple Cassini-Huygens flybys, with an approximate value of 3.1$\times10^4$ g~s$^{-1}$ \citep{muller-wodarg_titan_2014}. The Jeans escape of atomic and molecular hydrogen calculated in our models varies by a factor of a few or an order of magnitude from these observed escape rates. To the best of our knowledge, the individual escape rates of hydrogen ions and atomic nitrogen have not yet been measured by any instrument.
 
 By fitting a diffusion model to observed \ce{H2} abundance, \citet{yelle_vertical_2006} determined a 
 a \ce{H2} diffusive flux of $5.8\pm0.1~\times10^9$ cm$^{-2}$~s$^{-1}$ at an altitude of 1200 km. The \citet{yelle_vertical_2006} model adopted a $K_E$ of 4$\times 10^9$ cm$^{-2}$~s$^{-1}$ which is almost an order of magnitude higher than our Model B. The authors used their diffusion flux to calculate a diffusion-limited escape flux scaled to the surface (hereafter, in this sub-section, scaled to the surface flux is simply referred to as flux), which is equal to $1.2\pm0.2\times10^{10}$ cm$^{-2}$~s$^{-1}$. Using a larger sample of Cassini-Huygens' INMS data, \citet{cui_distribution_2008} also fitted a diffusive flux model to the observed \ce{H2} distribution and estimated a flux of 1.37 $\pm 0.01 \times 10^{10}$ cm$^{-2}$~s$^{-1}$. The eddy diffusion coefficient used in \citet{cui_distribution_2008} is based on \citet{yelle_methane_2008}. The diffusive flux in both studies is almost an order of magnitude higher than their estimated Jeans escape flux. \citet{cui_distribution_2008} state that the disagreement between Titan's Jeans escape and diffusion-limited escape of \ce{H2} can be addressed by adopting a non-Maxwellian velocity distribution for the gas at the exobase.

Instead of fitting our models to observations, we self-consistently solve for the temperature and chemical structure of Titan's upper atmosphere and calculate the thermal Jeans escape rate, which we then convert to escape flux for comparison against the diffusion-limited escape flux in the literature.
The calculation of Jeans escape reported in our models is based on \citet{bauer_planetary_2004} which considers a Maxwellian velocity distribution. As shown in Table~\ref{tab:escape_rates}, the Jeans escape flux of our models differ from the escape flux reported in \citet{yelle_vertical_2006} and \citet{cui_distribution_2008} by anywhere from a factor of a few to an order of magnitude, depending on the $K_E$ assumption in our model. This discrepancy in escape flux is within the expected uncertainties arising from differences in chemical networks, diffusion, and energy balance. Hence, we argue that a Maxwellian assumption is reasonable in the calculation of Jeans escape of \ce{H2} in Titan.

\begin{figure}
    \centering
    \includegraphics[width=0.45\textwidth]{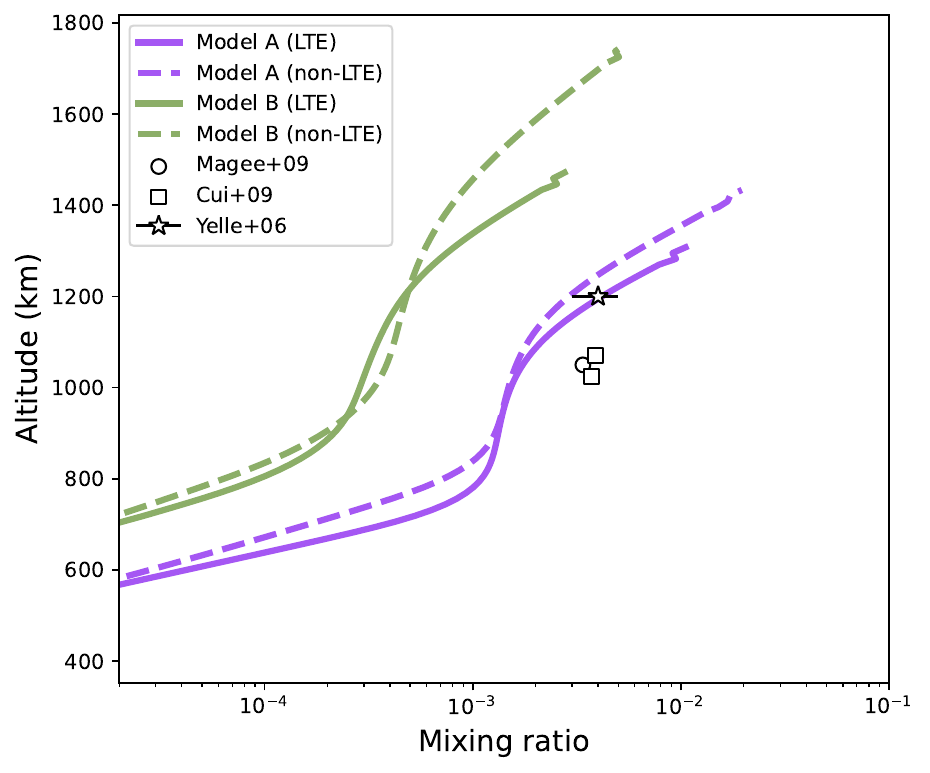}
    \caption{\ce{H2} mixing ratio vs altitude in km. The coloured lines represent the model results which follow the same colour scheme as in Fig.~\ref{fig:diffusion}. The markers represent \ce{H2} mixing ratios determined from Cassini-Huygens/INMS observations.}
    \label{fig:H2vmr}
\end{figure}

To further support our argument, we have plotted the \ce{H2} mixing ratio from our models together with Cassini-Huygens INMS derived \ce{H2} abundances in Fig.~
\ref{fig:H2vmr}. 
The observed values reported by \citet{magee_inms-derived_2009} and \citet{cui_analysis_2009} at an altitude of $\sim$1000 km are higher by a factor of 2-3 when compared against the LTE and the non-LTE Model A, and approximately an order of magnitude when compared against the LTE and the non-LTE Model B. The reported \ce{H2} abundances in \citet{yelle_vertical_2006} are in excellent agreement with Model A (LTE and non-LTE). The \citet{yelle_vertical_2006} results, however, disagree with Model B (LTE and non-LTE) by at least an order of magnitude. In general, our self-consistent models with weaker eddy diffusion coefficients are in better agreement with observations, and no strong preference is seen for either LTE or non-LTE models in such cases.

\begin{table*}
    \centering
    \caption{Top five chemical species by Jeans escape. Eight chemical species are shown, as the top five species are not the same in all models. Both escape rate in units of g s$^{-1}$ and flux in units of cm$^{-2}$s$^{-1}$ (in brackets) are included. Observationally determined escape rates and escape fluxes (in brackets) are also shown, separated from our model results by a horizontal line.}
    \begin{tabular}{ccccccccc}
    \hline
    \hline
        Model & \ce{H} & \ce{H2} & \ce{H+} & \ce{H2+} & \ce{H3+}  & \ce{CH4} & \ce {CH3}&\ce{N}\\
     & $\times 10^{5}$& $\times 10^{4}$($\times 10^{9}$) & $\times 10^{1}$ & $\times 10^{-4}$ & $\times 10^{-4}$ & $\times 10^{-3}$ & $\times 10^{-4}$ &$\times 10^{-2}$\\
     \hline
      A (LTE)  & 0.2   & 0.8(2.8) & 0.3  & 0.6  & 0.2& &  &\\
      B (LTE) &  0.6  & 0.4(1.4) & 0.2&& & 0.2(9.0)  & 0.6&\\
      A (non-LTE)&0.3&0.3(1.1)&0.4&2.9&&0.1(5.0)&&\\
      B (non-LTE)&0.1&1.4(5.1)&0.2&&&69.3(3.1$\times 10^3$)&&0.9\\
\hline
\citealt{hedelt_titans_2010} & 0.03 & & &&&&&\\
\citealt{muller-wodarg_titans_2014} & 0.02 & 3.1 & &&&&&\\
\citealt{yelle_vertical_2006} & & (12$\pm$2) & && &&\\
\citealt{cui_distribution_2008} && (13.7$\pm$0.1)&&&&&\\
\citealt{yelle_methane_2008} &&&&&&($2.8\times10^9$)&&\\
\citealt{cui_ch4_2012} &&&&&&($3.6\times10^9$)&&\\
\hline
    \end{tabular}
    \label{tab:escape_rates}
\end{table*}

\begin{figure}
    \centering
    \includegraphics[width=0.45\textwidth]{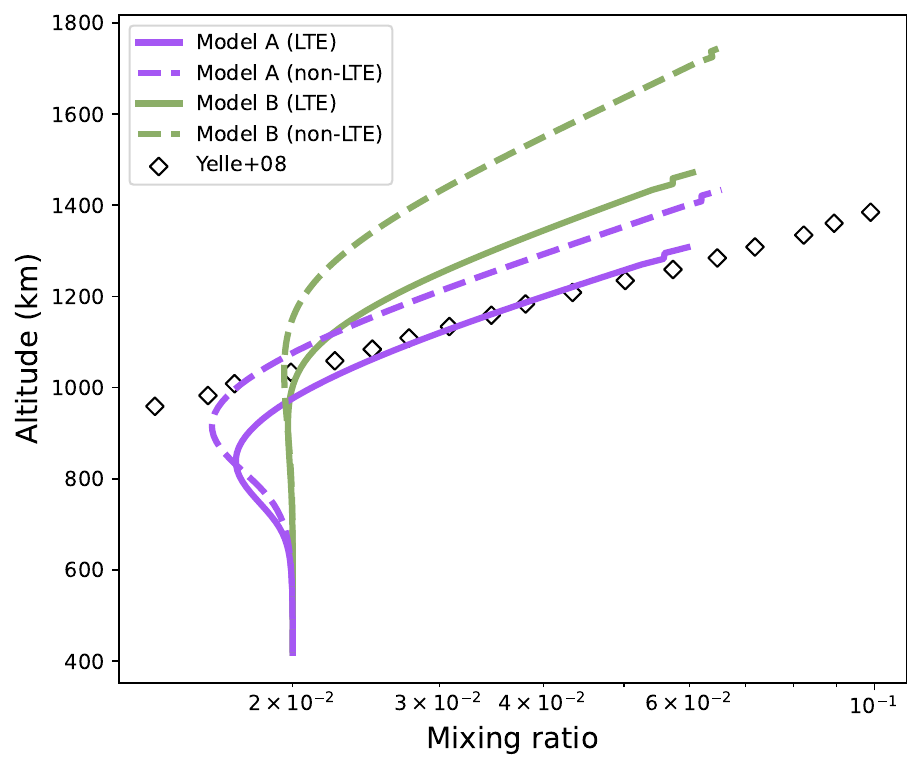}
    \caption{\ce{CH4} mixing ratio vs altitude in km, same colour scheme as Fig.~\ref{fig:diffusion}. The Cassini-INMS data analysed by \citet{yelle_methane_2008} are shown by the diamond markers.}
    \label{fig:ch4vmr}
\end{figure}

In addition to \ce{H2}, \ce{CH4} escape is speculated to play an important role in the long-term evolution of Titan's atmospheric chemistry and aerosol properties. However, there is no clear consensus around Titan's \ce{CH4} escape rate. \citet{yelle_methane_2008} hypothesises that a diffusive model with strong upward flow can explain high \ce{CH4} abundances, which can be as high as 10$\%$ in terms of the total gas near the exobase (Fig.~\ref{fig:ch4vmr}), and a \ce{CH4} escape flux of 2.8$\times$10$^9$ cm$^{-2}$~s$^1$ best fits this data. A newer study with additional Cassini-Huygens INMS data estimates the escape rate to be approximately 8$\times 10^4$ g~s$^{-1}$ \citep{cui_ch4_2012}, which leads to an escape flux of 3.6$\times 10^9$ cm$^{-2}$~s$^{-1}$. The Jeans escape of \ce{CH4} in our models is negligible. We calculated a \ce{CH4} escape flux of 9.0 cm$^{-2}$s$^{-1}$, 5.0 cm$^{-2}$s$^{-1}$ and 3.1$\times 10^3$ cm$^{-2}$s$^{-1}$ for Model B (LTE), Model A (non-LTE), and Model B (non-LTE) respectively, which are orders of magnitude lower than the estimated diffusion-limited escape fluxes.

Despite this large discrepancy in escape fluxes, the \ce{CH4} abundances in our models are in good agreement with \ce{CH4} abundances derived from Cassini-Huygens data. Fig.~\ref{fig:ch4vmr} shows that the \ce{CH4} abundances derived from Model A (LTE and non-LTE) show excellent agreement with the observed high abundance of \ce{CH4}, despite exhibiting negligible Jeans escape of \ce{CH4}. Model A (non-LTE) is at most a factor of two lower in abundance when compared to the observed data. At an altitude of 1000 km, the non-LTE Model A has marginally better agreement over the LTE Model A. The \ce{CH4} mixing ratio profiles dervied from the non-LTE Model B, our model with the largest \ce{CH4} escape flux, is lower than the observed abundances by an order of magnitude or more. The LTE Model B exhibits slightly lower abundances when compared against the non-LTE Model A. This shows that our self-consistent models reproduce observed abundances of \ce{CH4} without invoking strong eddy diffusion. 
The agreement between the observed and the model abundances in Fig.~\ref{fig:ch4vmr} do not support the theory of strong atmospheric \ce{CH4} escape into space.

\subsection{Temporal evolution of Titan's upper atmosphere}
Although our models predict negligible thermal escape of \ce{CH4} under present-day Titan conditions, we investigate whether \ce{CH4} could have been lost more efficiently through thermal Jeans escape earlier in Titan's history. Since the abundance of atmospheric \ce{CH4} over Titan's history is not well known, we retain the present-day atmospheric composition in all simulations. It has been suggested that the atmospheric \ce{CH4} reservoir on present-day Titan is relatively young, with an age of $\leq$ 1 Ga \citep{tobie_episodic_2006}. Therefore, \ce{CH4} may not have been present in significant amounts 2--3 Ga ago. Nevertheless, we extend our simulations to 3 Ga as a hypothetical exercise to investigate how Titan's upper atmosphere would have responded to the enhanced solar XUV flux if Titan's present-day atmosphere had existed at those epochs. 

\begin{figure}
    \centering
    \includegraphics[width=0.5\textwidth]{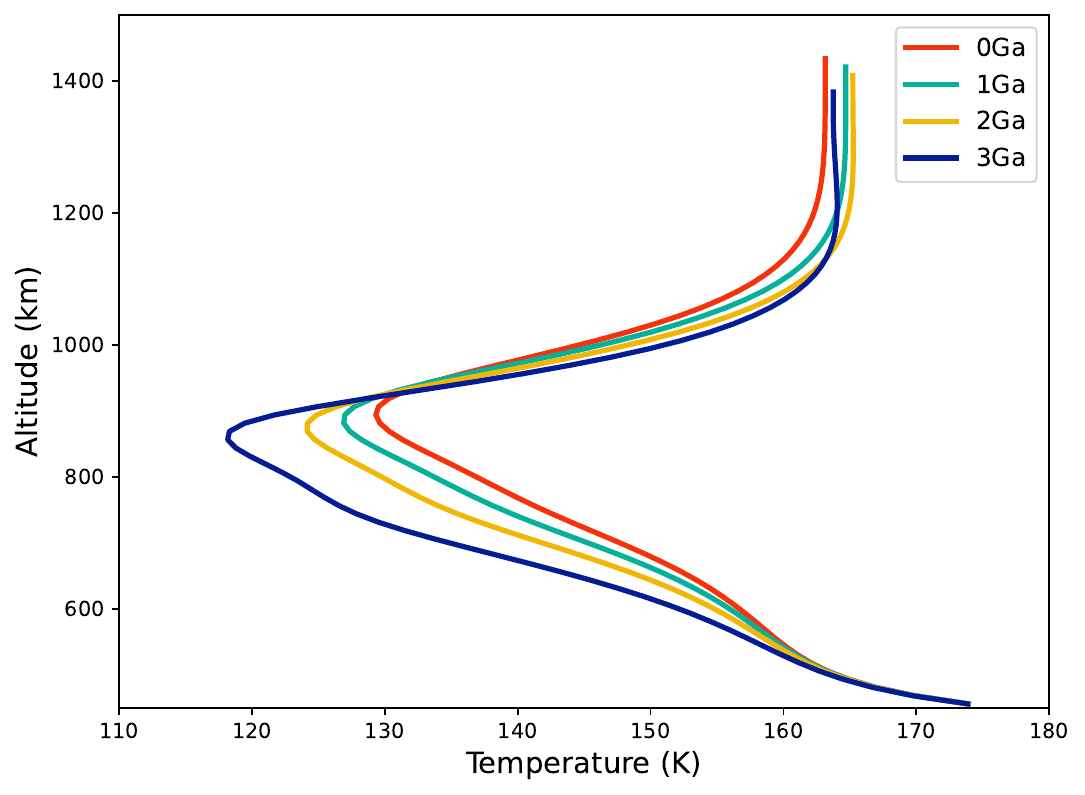}
    \caption{Titan's upper atmospheric temperature over time. The  orange line represents present-day Titan, specifically Model A (non-LTE), the green, yellow and dark blue lines represent 1 Ga, 2 Ga, and 3 Ga in the past respectively.}
    \label{fig:placeholder}
    \label{fig:thermal_time}
\end{figure}

\begin{figure}
    \centering
    \includegraphics[width=0.5\textwidth]{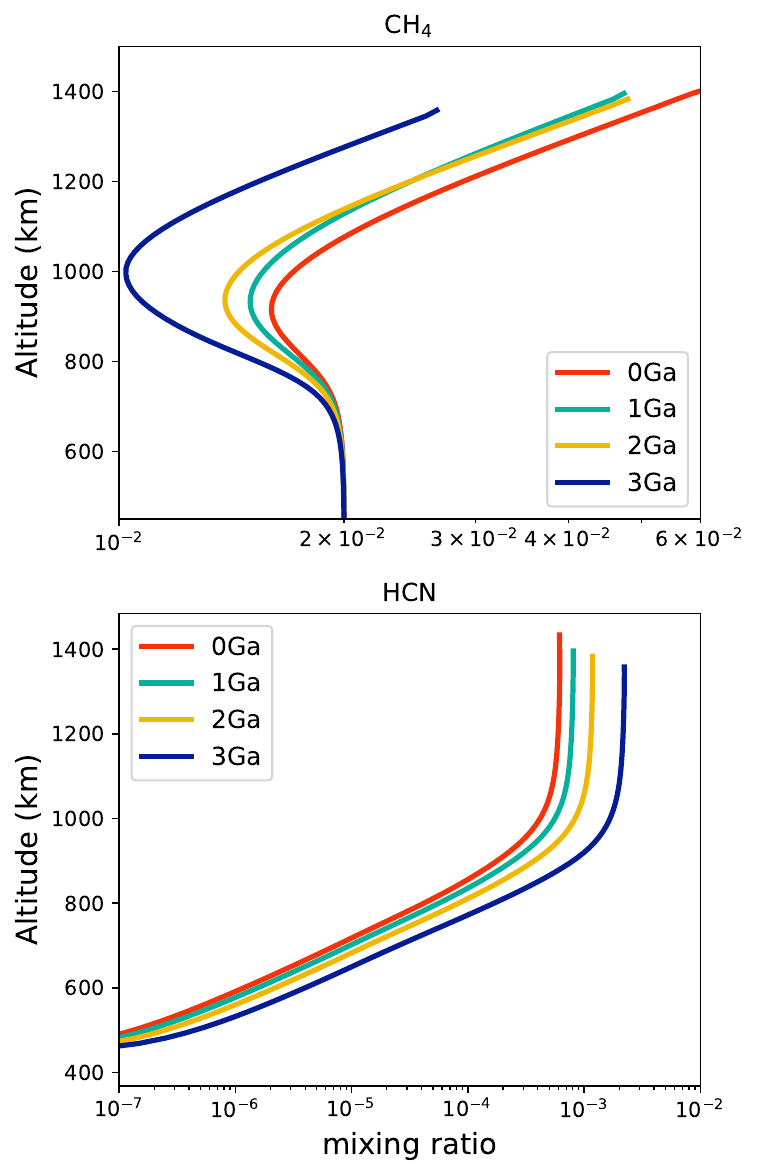}
    \caption{Mixing ratio of \ce{CH4} ({\it Top}) and \ce{HCN} ({\it Bottom}) over time. Same colour scheme as Figure~\ref{fig:thermal_time}}
\label{fig:species_time}
\end{figure}

 We explore Titan's atmospheric evolution using the same set of assumptions as used in Model A (non-LTE), as it offers the best observationally constrained representation of Titan's upper atmosphere. We use solar XUV spectra in time generated by the model of \citet{claire_evolution_2012}. Figure \ref{fig:thermal_time} shows the temporal evolution of Titan's upper atmospheric temperature structure, which doesn't show any significant increase in the upper thermosphere, but instead exhibits a relatively cooler temperature in the middle thermosphere for early Titan. 
 The increase in solar XUV flux between now and 3 Ga ago leads to increased XUV absorption at the upper atmosphere. Although the solar XUV flux is stronger in the past, much of the XUV energy is absorbed in the upper thermosphere, limiting the increase in XUV heating at lower altitudes (see Appendix \ref{energy_time} for further details). In addition, an increase in the XUV flux also enhances the production of \ce{HCN}, via the photolysis of \ce{CH4} and \ce{N2}. Figure \ref{fig:species_time} shows the temporal evolution of \ce{HCN} and \ce{CH4} abundance in Titan's upper atmosphere. The enhanced \ce{HCN} abundance increases the \ce{HCN} number density throughout the upper atmosphere, strengthening \ce{HCN} radiative cooling.  Since the enhanced \ce{HCN} cooling is not fully compensated by local XUV heating, it results in lower temperatures in the middle thermosphere.

The Jeans escape from our three additional steady-state models exhibit a similar trend as the 0 Ga model (non-LTE Model A). Atomic hydrogen (\ce{H}) exhibits the strongest Jeans escape in all three models. Jeans escape of \ce{CH4} is identical between 0 Ga and 1 Ga. However, at 2 and 3 Ga, Jeans escape of \ce{H}, \ce{H2} and {H} ions dominate over \ce{CH4} loss. This is not surprising, as the abundance of \ce{CH4} decreases with increase in time from the present. 

\section{Conclusions}

We updated the 1D atmospheric code Kompot, and benchmarked our model against observations of Titan's upper atmospheric thermal and chemical structure. We further assessed the sensitivity of the model to the choice of eddy diffusion coefficient profile and to the treatment of \ce{CH4} radiative cooling under LTE and non-LTE assumptions. 

Our simulated upper atmospheric temperature structures fall within the range of temperature profiles derived from observations. This suggests that energy balance between \ce{HCN} and \ce{CH4} radiative cooling, XUV and chemical heating, and conduction reproduces Titan's thermospheric temperature profile within observed variations. Additional physics not considered in this work, such as aerosol heating, could modify the temperature structure locally, but the overall temperature structure would still be governed by the key mechanisms included in our model. The non-LTE Model A and the LTE Model B produce temperature profiles that both fall within the central portion of the observed variability of Titan's upper atmospheric temperature measurements. Although the two models differ in their assumptions regarding eddy diffusion and \ce{CH4} radiative cooling, they achieve comparable agreement with the observations. This suggests that the thermal structure alone does not uniquely constrain these model parameters, as different combinations of eddy diffusion and radiative cooling treatment can reproduce similar temperature profiles.

The chemical abundance profiles from our models show varying levels of agreement with observed abundances. The model \ce{N2} and \ce{CH4} number density profiles exhibit a similar trend as the temperature structure, where Models A (non-LTE) and Model B (LTE) show the strongest agreement with observational data. However, for the minor species, the level of agreement between our models and observations is species dependent. Furthermore, depending on the observational dataset, either Models B (LTE and non-LTE) or Models A (LTE and non-LTE) are in a better agreement with the data. Comparison of our model \ce{Ar} abundance profile show that both LTE and non-LTE Model A are broadly consistent with the observations, but Model A (non-LTE) provide a better match with observations. Since \ce{Ar} is a noble gas, its abundance profile is governed primarily by vertical transport, providing a better constraint for the eddy diffusion profile. Thus, when combined with the temperature structure results, the non-LTE Model A, with weaker eddy diffusion, provides the most consistent description of Titan's upper atmospheric temperature structure and chemistry.

Our results highlight the importance of self-consistently modelling the coupled thermal and chemical structure of planetary upper atmospheres. This is particularly relevant for exoplanet atmospheres, where the temperature structure is not known a priori. Different transport and radiative processes can produce similar temperature profiles, while their coupled effects on the chemical composition and temperature can differ substantially. Simultaneously modelling these feedbacks, and combining thermal and chemical constraints, can therefore help break certain degeneracies and constrain the underlying processes in planetary upper atmospheres.

We also calculated the Jeans escape rates at Titan's exobase and identified neutral and ionised hydrogen to have the largest escape fluxes. Nonetheless, our thermal \ce{H2} escape fluxes are weaker than those inferred in the literature under the diffusion-limited regime, where diffusion models are fitted to observationally derived \ce{H2} abundances. While comparing our model \ce{H2} abundances with the observed values, we find that models with weaker eddy diffusion coefficients, both LTE and non-LTE Model A, exhibit better agreement with the data. We hypothesize that the external influx of \ce{H2O} from Enceladus, rather than very strong diffusion, could at least partially explain the observed overabundance of \ce{H2} in Titan's upper thermosphere.
The Jeans escape of \ce{CH4} is negligible in our models, except in the case of Model B (non-LTE). However, the \ce{CH4} escape flux in Model B (non-LTE) is more than six orders of magnitude lower when compared to diffusion-limited escape flux calculations from Cassini-Huygens observations. Our models show that a relatively high abundance of \ce{CH4}, similar to the ones seen in Cassini-Huygens observations, can be achieved without invoking strong diffusion or escape rates. Our results do not support strong \ce{CH4} escape from Titan's atmosphere, even when early Sun's XUV fluxes are considered.

Finally, our model is one-dimensional, whereas Titan’s atmosphere is three-dimensional, and the available observations span different seasons, local times, and atmospheric pressure regimes. Future work will therefore require both increasingly comprehensive models and observations that provide more coordinated constraints on the atmospheric thermal and chemical structure. As an example, inclusion of energetic and radiative feedback from haze particles in the model would enhance 
   our predictive capability for the mesospheric thermal and chemical structure. As model complexity increases, coordinated observations that constrain the same atmospheric regions and conditions will be essential for distinguishing between the effects of different atmospheric processes and for ensuring that model complexity
remains physically and observationally justified. These considerations are also directly relevant to exoplanet atmospheric modelling, where observational constraints are typically much more limited and Titan provides an important Solar-system laboratory for developing and testing such approaches.
\section*{Acknowledgements}

We thank the anonymous referee for their valuable comments which have helped us improve the manuscript. This work was partly supported by the ASAP project APPLE-Kompot, funded by the Austrian Forschungsf\"orderungsgesellschaft (FFG) through BMIMI, and the European Research Council (ERC) project EASE, 101123041. Work performed at UCL was supported by the ERC under the
European Union’s Horizon 2020 research and innovation programme via Advanced Grant 883830. Views and opinions expressed are however those of the author(s) only and do not necessarily reflect those of the European Union or the European Research Council Executive Agency. Neither the European Union nor the granting authority can be held responsible for them.

\section*{Data Availability}
The observational data used in this work are taken from previously published work. The data related to the chemical network update is provided in the appendix.



\bibliographystyle{mnras}
\bibliography{references} 




\appendix

 \section{Heating and cooling rates for present-day Titan}~\label{sec:heatcool}
\begin{figure}
         \centering
         \includegraphics[width=0.5\textwidth]{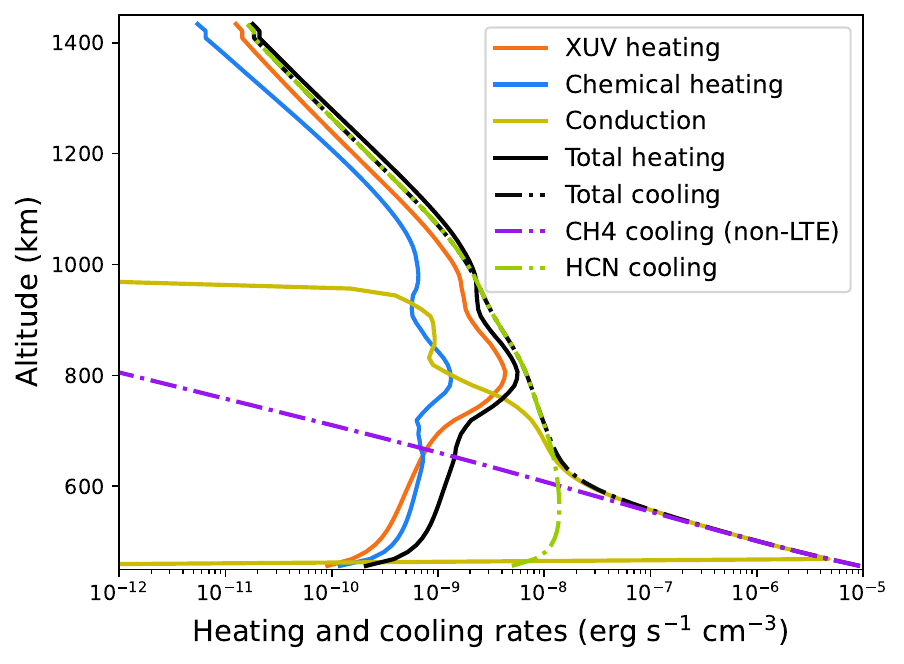}
         \caption{Total heating and cooling rates for Model A (non-LTE). The dominant heating and cooling functions are also shown by the coloured lines.}
         \label{fig:total_energy}
     \end{figure}

     {Figure \ref{fig:total_energy} shows the key contributors to the total heating and cooling rate for Model A (non-LTE). The total heating rate is dominated by XUV heating, followed by chemical heating. Infrared heating is the third dominant heating mechanism, but it is orders of magnitude weaker than XUV and chemical heating. At the lower thermosphere the total cooling is dominated by radiative cooling by \ce{CH4}, but \ce{HCN} cooling takes over at the middle and upper thermosphere. Near the lower boundary of our simulations, thermal conduction balances the strong radiative cooling rate by \ce{CH4}. The energy balance between these heating and cooling profiles, together with conduction, leads to the temperature structure in our four atmospheric models, as shown in Fig.~
 \ref{fig:PTprofile}.}
 
\begin{figure}
    \centering

    \includegraphics[width=0.5\textwidth]{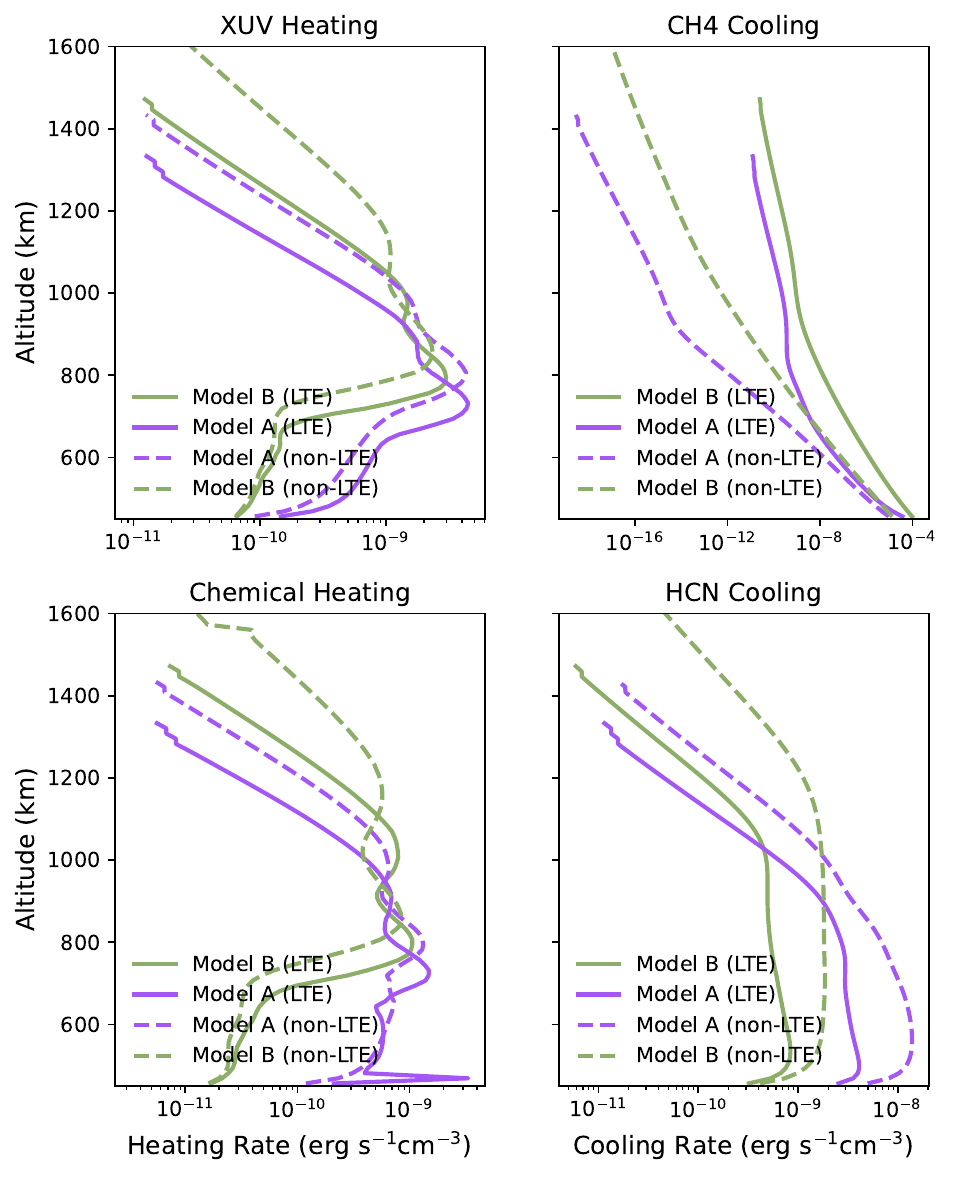}
    \caption{Heating ({\it left}) and cooling ({\it right}) rates in erg s$^{-1}$ cm$^{-3}$ vs altitude (km). The colour scheme is the same as in Fig.~\ref{fig:diffusion} }
\label{fig:heating_cooling_vol}
\end{figure}
\begin{figure}
    \centering
    \includegraphics[width=0.5\textwidth]{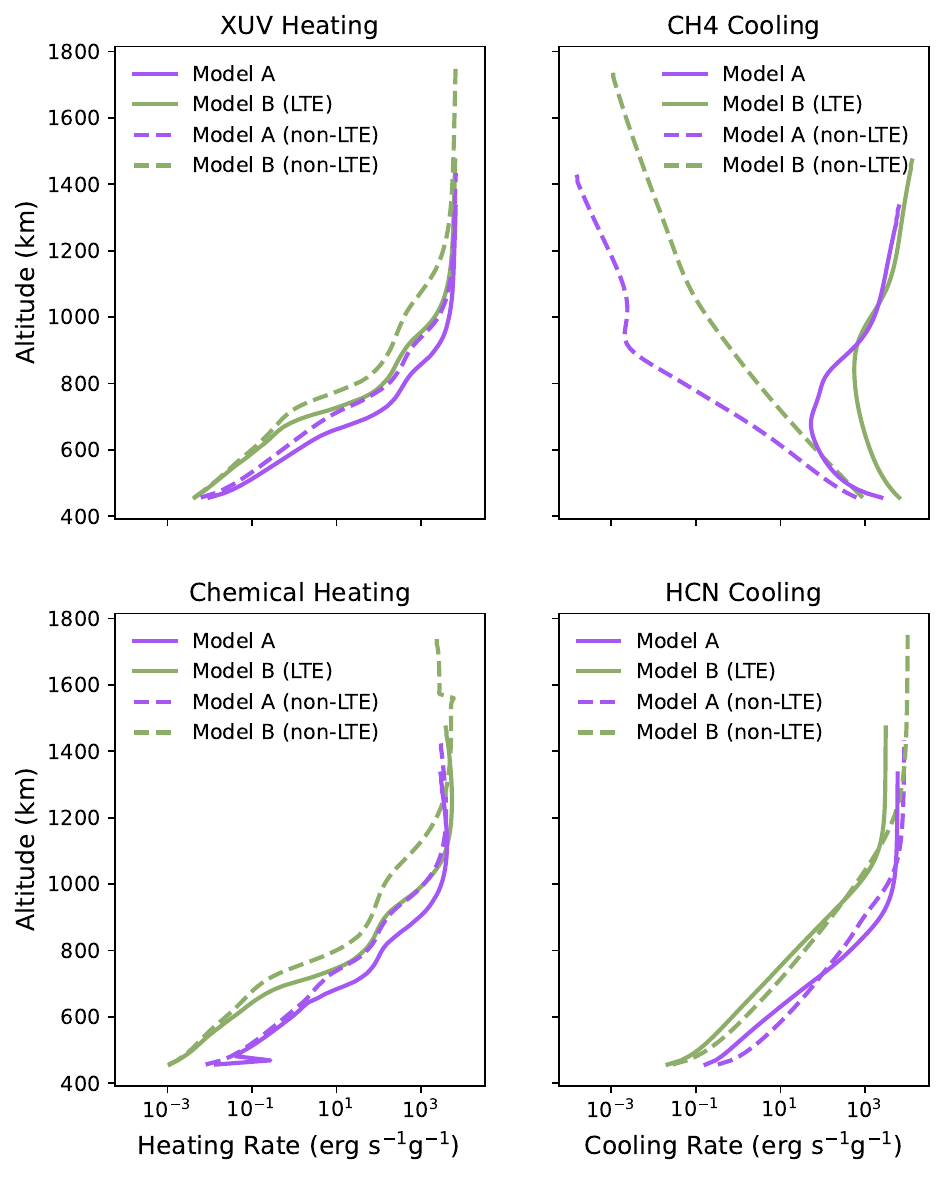}
    \caption{Heating ({\it left}) and cooling ({\it right}) rates in erg s$^{-1}$ g$^{-1}$ vs altitude (km). The colour scheme is the same as in Fig.~\ref{fig:diffusion} }
\label{fig:heating_cooling_mass}
\end{figure}

Figures~\ref{fig:heating_cooling_vol}~and~\ref{fig:heating_cooling_mass} show the volumetric energy rates and the energy rates per unit mass for the dominant heating and cooling processes in our models. The volumetric XUV heating (Fig.~\ref{fig:heating_cooling_vol}, top left) peaks in the mid-to-lower thermosphere, as it relies on the density of the absorbing gas and the amount of XUV photons available at each altitude. Whereas the XUV heating rate per mass (Fig.~\ref{fig:heating_cooling_mass}, top left) is stronger in the upper thermosphere as it is normalised by the local density profile of the absorbing gas. While the characteristics of the XUV heating rate are identical across the  four models, the absolute values of the heating rate are different between models. Models with a stronger eddy diffusion coefficient show a weaker XUV heating rate per unit mass, but the volumetric heating rate shows a more complex behaviour. At the upper thermosphere, the heating rate per unit volume is higher for models with stronger eddy diffusion coefficients, which can be attributed to the more efficient transport of heavier XUV-active molecules to higher altitudes. {The volumetric XUV heating rate is also higher for the non-LTE models, which could be due to less effective cooling in these models leading to be a more expanded atmosphere.} Both phenomena could also explain the opposite trend in the volumetric heating rate pxrofile at the lower thermosphere. 

The chemical heating (bottom left panels in Fig.~\ref{fig:heating_cooling_vol} and~\ref{fig:heating_cooling_mass}) exhibits broadly similar behaviour to the XUV heating rate. Differences in the vertical distribution of chemical species between the four models could explain observed inter-model variations in the chemical heating rate. Our model treats the stellar IR spectrum as a blackbody of 5777 K and currently only considers IR absorption by \ce{CO2} and \ce{H2O}. Since the abundance of these two absorbing molecules is negligible  in Titan's thermosphere and also in our models, heating due to IR absorption is negligible. Both \citet{yelle_non-lte_1991} and \citet{snowden_thermal_2014} accounted for heating due to IR absorption in the \ce{CH4} bands.  A detailed investigation of the impact of IR absorption by \ce{CH4} is beyond the scope of this work and will be left for the future, but its contribution towards the energy calculations is expected to be minor compared to XUV heating.

The total cooling rate in all four models is driven by the thermal emission of \ce{CH4} and \ce{HCN}. 
{The volumetric cooling rates under the LTE assumption are derived from the cooling rate per molecule and the density profile of that molecule. As shown in the top right panel of Fig.~\ref{fig:heating_cooling_vol}, the \ce{CH4} volumetric cooling rates for our LTE models decrease in strength from Model B to A, showing a correlation with the adopted $K_E$ profile. 
In Fig.~\ref{fig:heating_cooling_mass}, the \ce{CH4} cooling per mass for the LTE models at the upper thermosphere doesn't exhibit any dependence on $K_E$. At the lower thermosphere, however, the \ce{CH4} cooling rate per mass increases with an increase with $K_E$. 
The non-LTE models by construction have a weaker \ce{CH4} cooling when compared to the LTE models, as departures from LTE reduces cooling efficiency. Between the two non-LTE models, \ce{CH4} cooling is stronger for the model with a higher $K_E$.}

{The \ce{HCN} cooling rate is calculated under the LTE assumption only. Any reference to a non-LTE model point towards non-LTE treatment of \ce{CH4} cooling in that model. The bottom right panels in Fig.~\ref{fig:heating_cooling_vol} and ~\ref{fig:heating_cooling_mass} shows the \ce{HCN} cooling rates. The volumetric \ce{HCN} cooling exhibits a more complex behaviour when compared to the \ce{HCN} cooling per unit mass which shows a consistently weaker \ce{HCN} cooling profile with increase in $K_E$. At the lower thermosphere the volumteric \ce{HCN} cooling rate is weaker for the stronger $K_E$ models, but at the upper thermosphere the cooling rate is increases for the stronger $K_E$ models. This is not surprising as \ce{HCN} is a product of \ce{CH4} and \ce{N2} chemistry, and its density structure is strongly tied to the adopted $K_E$ profile which impacts the volumteric cooling rate of this molecule. Finally, the volumetric cooling rate of the non-LTE models is stronger than the LTE models, which could be explained by the overall hotter and expanded atmosphere due to a less efficient \ce{CH4} cooling in those models which impacts the production of \ce{HCN} and consequently the \ce{HCN} density in these models.} 

\section{Energy deposition in Titan's upper atmosphere over time}~\label{energy_time}
\begin{figure}
    \centering
    \includegraphics[width=0.5\textwidth]{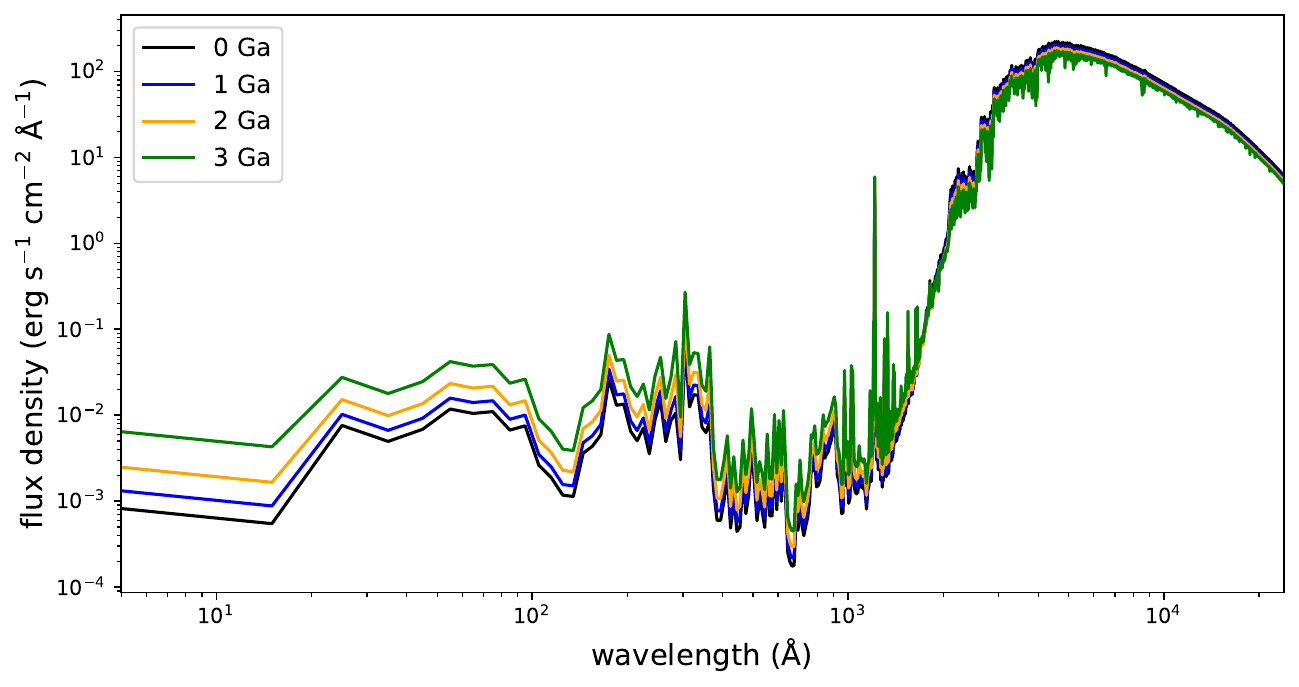}
    \caption{Solar X-ray and UV spectrum from the \citet{claire_evolution_2012} model over time, from 0 to 3 Ga in the past.}
    \label{fig:sun}
\end{figure}

\begin{figure}
    \centering
    \includegraphics[width=0.5\textwidth]{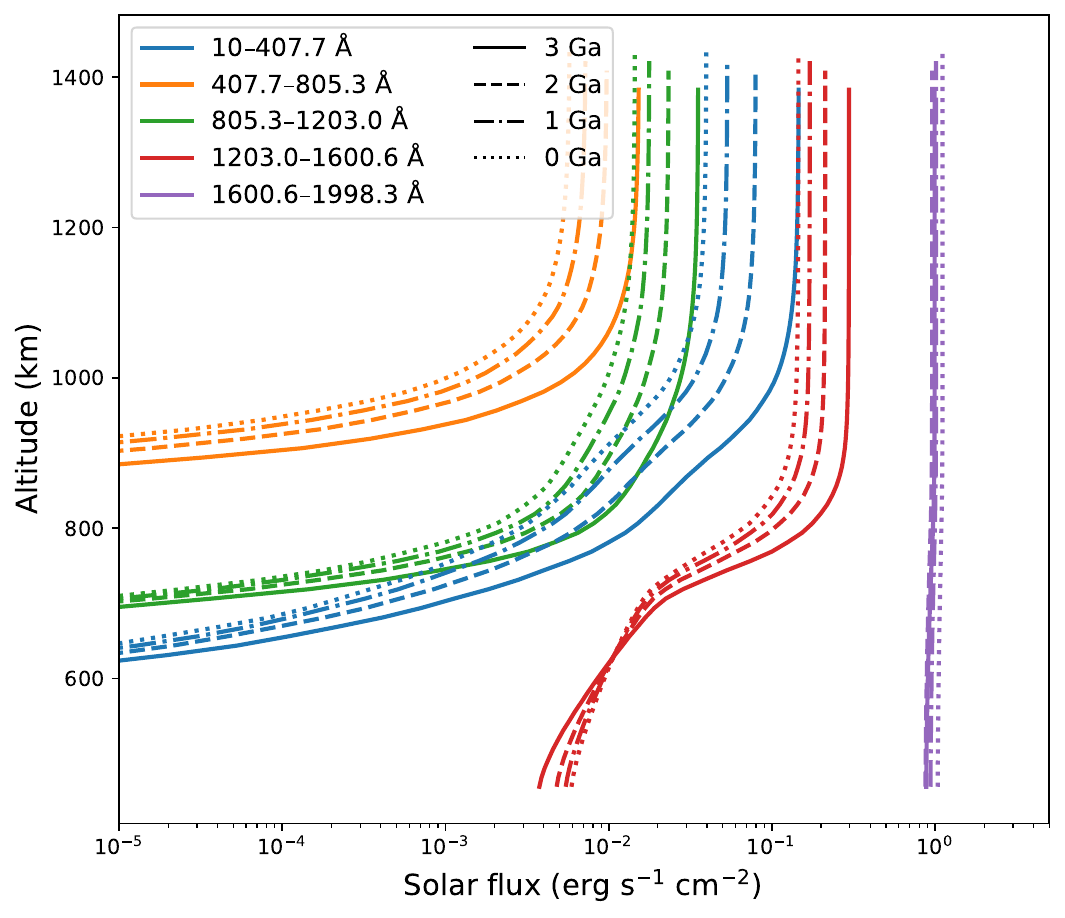}
    \caption{XUV flux transfer through Titan's upper atmospheric temporal models, from 0 to 3 Ga in the past (marked by the different line styles). The colours represent different wavelength bins, as listed in the legend.}
    \label{fig:absorption_xuv}
\end{figure}

Figure~\ref{fig:sun} shows the temporal evolution of the Solar XUV spectrum  over time, from 0 Ga (present) to 3 Ga in the past, based on the model of \citet{claire_evolution_2012}. At shorter wavelengths, the spectral irradiance of the young Sun is up to eight times higher than that of the present-day Sun, whereas this difference progressively decreases toward longer wavelengths. The attenuation of this solar XUV flux in Titan's upper atmosphere is strongly wavelength-dependent, with photons of different wavelengths being absorbed at different altitudes as the radiation propagates downward through the atmosphere, as illustrated in Fig.~\ref{fig:absorption_xuv}. While majority of the X-ray and far-UV flux gets absorbed above 800 km, a vast majority of the EUV flux gets absorbed above 1000 km. In contrast, the near-UV  radiation at wavelengths longer than 1600 \AA~ experiences negligible absorbtion in the upper atmosphere. Overall, the transmitted flux is highest for the youngest Sun, consistent with its enhanced incident XUV irradiance in Fig~\ref{fig:sun}. However, this trend shows a weak reversal in the near-UV, where the present-day Sun produces a slightly higher transmitted flux because its incident spectral irradiance exceeds that of the younger Sun at these wavelengths.

\onecolumn
\begin{table*}[!ht]
\section{Chemical reactions}\label{chem_appendix}

\centering
\caption{Photoreactions}
\label{table:photoreaction}
\begin{threeparttable}
    
\begin{tabular}{ccc}
\hline
\hline
{No.} & {Reaction} & {Ref.} \\
\hline
R1 & \ce{CH4 + {$\mathit{h}\nu$} -> CH3 + H} & 1\\
R2  & \ce{CH4 + {$\mathit{h}\nu$} -> ^1CH2 + H2} & 1\\
R3  & \ce{CH4 + {$\mathit{h}\nu$} -> CH2 + H2} & 1\\
R4  & \ce{CH4 + {$\mathit{h}\nu$} -> CH + H2 + H} & 1 \\
R5 & \ce{C2H2 + {$\mathit{h}\nu$} -> C2H + H} &  1\\
R6  & \ce{C2H2 + {$\mathit{h}\nu$} -> C2 + H2} &  1\\
R7 & \ce{C2H4 + {$\mathit{h}\nu$} -> C2H2 + H2} &  1\\
R8 & \ce{C2H4 + {$\mathit{h}\nu$} -> C2H2 + H + H} & 1 \\
R9 & \ce{C2H6 + {$\mathit{h}\nu$} -> C2H4 + H2} &  1\\
R10 & \ce{C2H6 + {$\mathit{h}\nu$} -> CH4 + ^1CH2} & 1 \\
R11 & \ce{C2H6 + {$\mathit{h}\nu$} -> CH3 + CH3} & 1 \\
R12 & \ce{C4H2 + {$\mathit{h}\nu$} -> C4H + H} &  2\\
R13 & \ce{C4H2 + {$\mathit{h}\nu$} -> C2H + C2H} & 2\\
R14 & \ce{C4H4 + {$\mathit{h}\nu$} -> C4H2 + H2} & 2 \\
R15 & \ce{NH3 + {$\mathit{h}\nu$} -> NH2 + H} & 1\\
R16 & \ce{NH3 + {$\mathit{h}\nu$} -> NH + H + H}& 1\\
R17  & \ce{HCN + {$\mathit{h}\nu$} -> H + CN} & 1 \\
R18  & \ce{HC3N + {$\mathit{h}\nu$} -> C2H + CN} & 1 \\
R19  & \ce{C2N2 + {$\mathit{h}\nu$} -> CN + CN} & 2 \\
\hline
\end{tabular}
\begin{tablenotes}
\footnotesize
\item[1] \citet{huebner_photoionization_2015}
\item[2] \citet{keller-rudek_mpi-mainz_2013}
\end{tablenotes}
\end{threeparttable}
\end{table*}%
\FloatBarrier

\begin{longtable}{cccccc}
\caption{Thermal reactions involving neutrals} 
\label{table:chem_neutral}\\
\hline
No. & Reaction & Energy & Temperature & Rate constant ($k_0$) & Ref. \\
&&eV&K&&\\
\hline
\endfirsthead
\caption{Thermal reactions involving neutrals} \\
\hline
No. & Reaction & Energy & Temperature & Rate constant($k_0$) & Ref. \\
&&eV&K&&\\

\hline
\endhead

\hline
\endfoot

\hline
\endlastfoot


R1 & \ce{CH + CH4 -> C2H4 + H} & 2.5 & 10--298 & $1.6 \times 10^{-10} \left( \frac{T}{300} \right)^{-1.04} \exp\left(\frac{-36.1}{T}\right)$ \\
R2 & \ce{CH + CH4 -> CH2 + CH3} & -0.2 & 10--800 & $2.28 \times 10^{-11} \left( \frac{T}{300} \right)^{0.7} \exp\left(\frac{-3000}{T}\right)$ & 1 \\ 
R3& \ce{^1CH2 + CH4 -> CH3 + CH3} & 0.65 & 43--498& $7.87 \times 10^{-11} \left( \frac{T}{300} \right)^{-0.467} \exp\left(\frac{-28.8}{T}\right) $ & 2\\ 
R4&\ce{^1CH2 + C2H4 -> H + C3H5} &-- &43--498 & $1.88\times10^{-9} \left(\frac{T}{300}\right)^{-3.56} \exp\left(\frac{-724}{T}\right)$ & \\
R5&\ce{^1CH2 + C2H5 -> H + C3H6} &--  &50--200 & $1.50\times10^{-11}$ &\\
R6 & \ce{^1CH2 + N2 -> CH2 + N2} & 0.4 & 43--200 & $6.00 \times 10^{-12} \left( \frac{T}{300} \right)^{-0.739}$ & 3 \\
R7 & \ce{^1CH2 + N2 -> CH2 + N2} & 0.4 & 200--798 & $6.55 \times 10^{-12} \left( \frac{T}{300} \right)^{1.17} \exp\left(\frac{-153}{T}\right)$ & 4\\
R8 & \ce{^1CH2 + H2 -> CH3 + H} & 0.3 & 10--280 & $5.0 \times 10^{-11} \exp\left(\frac{-4870}{T}\right)$ \\

R9 & \ce{CH2 + CH2 -> C2H2 + H2} & 1.7 & 50--200 & $5.3 \times 10^{-11}$ & 5\\
R10 & \ce{CH2 + CH3 -> C2H4 + H} & 2.7 & 50--200 & $7.0 \times 10^{-11}$ & 2\\
R11 & \ce{CH2 + CH3 -> C2H4 + H} & 2.7 & 298--3000 & $1.2 \times 10^{-10}$ & 6 \\ 
R12 & \ce{CH3 + CH3 + M -> C2H6 + M} & 3.8 & 50--200 & $1.13 \times 10^{-25} \left( \frac{T}{300} \right)^{-3.75} \exp\left(\frac{-494}{T}\right)$ & 2\\
R13 & \ce{CH3 + CH3 + M -> C2H6 + M} & 3.8 & 300--2000 & $1.6 \times 10^{-24} \left( \frac{T}{300} \right)^{-7.0} \exp\left(\frac{-1390}{T}\right)$ & 6\\
R14&\ce{CH3 + C2H3 -> C2H2 + CH4} &-- &10--800 & $3.30\times10^{-11}$ &\\
R15&\ce{CH3 + C2H4 -> C2H3 + CH4} &-- &50--200 & $1.61\times10^{-14} \left(\frac{T}{300}\right)^{3.7} \exp\left(\frac{-4780}{T}\right)$ &\\
R16&\ce{CH3 + C2H5 -> C2H4 + CH4} &-- &50--200 & $1.90\times10^{-12}$ &\\
R17&\ce{CH3 + C2H5 -> C2H4 + CH4} &--&300--2000 & $1.50\times10^{-12}$& \\
R18&\ce{CH3 + C2H6 -> C2H5 + CH4} &-- &50--200 & $1.82\times10^{-16} \left(\frac{T}{300}\right)^6 \exp\left(\frac{-3040}{T}\right)$ &\\
R19 & \ce{CH3 + C2H5 + M -> C3H8 + M} & -- & 50--200 & $1.05 \times 10^{-19} \left( \frac{T}{300} \right)^{-16.1} \exp\left(\frac{-1900}{T}\right) $ \\
R20&\ce{CH3 + C3H5 -> CH3C2H + CH4} &--& 50--200 & $4.03\times10^{-13} \left(\frac{T}{300}\right)^{-0.32} \exp\left(\frac{66}{T}\right)$ &\\
R21&\ce{CH3 + C3H8 -> C3H7 + CH4} &--& 50--200 & $1.65\times10^{-15} \left(\frac{T_{\text{neut}}}{300}\right)^{3.65} \exp\left(\frac{-3600}{T_{\text{neut}}}\right)$ &\\
R22&\ce{CH3 + C4H3 -> C4H2 + CH4} &-- &50--200 & $3.0\times10^{-11}$ &\\
R23&\ce{CH3 + C4H4 -> C4H3 + CH4} & --&50--200 & $6.61\times10^{-13} \exp\left(\frac{-2500}{T}\right)$ &\\
R24 & \ce{CH3 + N -> HCN + H2} & 5.07 & 10--300 & $5.00 \times 10^{-11} \exp\left(\frac{-250}{T}\right)$ &5 \\
R25 & \ce{CH3 + NH2 -> CH4 + NH} & 0.45 &-- & $4.65\times10^{-18}\left(\frac{T}{300}\right)^{1.94}\exp\left(\frac{-4635}{T}\right)$ & 7\\
R26&\ce{CH3 + NH2 -> NH3 + CH2} & -0.18 & -- & $2.66 \times 10^{-18} \left(\frac{T}{300}\right)^{1.87} \exp\left(-\frac{3810}{T}\right)$ & 7 \\
R27&\ce{CH3 + NH3 -> NH2 + CH4} &-- &50--200 & $5.10\times10^{-14} \left(\frac{T}{300}\right)^{2.86} \exp\left(\frac{-7340}{T}\right)$& \\
R28&\ce{CH3 + H2 -> CH4 + H} &-- &50--200 & $6.99\times10^{-14} \left(\frac{T}{300}\right)^{2.74} \exp\left(\frac{-4740}{T}\right)$ &\\
R29 & \ce{C2 + CH4 -> C2H + CH3} & 0.4 & 50--200 & $5.05 \times 10^{-11} \exp\left(\frac{-297}{T}\right)$ & 2\\
R30 & \ce{C2 + H2 -> C2H + H} & 1.5 & 50--200 & $1.77 \times 10^{-10} \exp\left(\frac{-1470}{T}\right)$ & 2 \\
R31 & \ce{C2H + CH4 -> C2H2 + CH3} & 1.2 & 150--780 & $7.67 \times 10^{-12} \left( \frac{T}{300} \right)^{0.94} \exp\left(\frac{-328}{T}\right)$ & 6 \\
R32 & \ce{C2H + C2H6 -> C2H2 + C2H5} & 1.3 & 150--780 & $3.33 \times 10^{-11} \left( \frac{T}{300} \right)^{-0.28} \exp\left(\frac{62}{T}\right)$ & 6\\
R33 & \ce{C2H + C2H2 -> C4H2 + H} & 1.3 & 50--2700 & $1.3 \times 10^{-10}$ & 2, 6\\
R34 & \ce{C2H + HCN -> HC3N + H} & 1.2 & 50--200 & $5.3 \times 10^{-12} \exp\left(\frac{-770}{T}\right)$ &2\\
R35 & \ce{C2H + H2 -> C2H2 + H} & 1.3 & 180--300 & $1.95 \times 10^{-12} \left( \frac{T}{300} \right)^{2.32} \exp\left(\frac{-444}{T}\right)$ & 6\\
R36 & \ce{C2H + H2 -> C2H2 + H} & 1.3 & 50--180 & $1.2 \times 10^{-11} \exp\left(\frac{-998}{T}\right)$ & 2\\
R37 & \ce{C2H2 + H + M -> C2H3 + M} & 1.5 & 50--200 & $3.3 \times 10^{-30} \exp\left(\frac{-740}{T}\right)$ & 2 \\
R38 & \ce{C2H3 + H -> C2H2 + H2} & 3.0 & 300 & $7.0 \times 10^{-11}$ & 6\\
R39 & \ce{C2H3 + H -> C2H2 + H2} & 3.0 & 50--200 & $2.0 \times 10^{-11}$ & 2\\
R40 & \ce{C2H3 + C2H3 -> C2H2 + C2H4} & 0.1 & 50--200 & $3.5 \times 10^{-11}$ & 2 \\
R41 & \ce{C2H4 + H + M -> C2H5 + M} & 1.4 & 50--200 & $1.39 \times 10^{-29} \exp\left(\frac{-562}{T}\right)$ & 2\\
R42 & \ce{C2H4 + H + M -> C2H5 + M} & 1.4 & 300--800 & $1.30 \times 10^{-29} \exp\left(\frac{-380}{T}\right)$ & 6\\
R43 & \ce{C2H5 + H -> C2H4 + H2} & 2.9 & 10--300 & $3.0 \times 10^{-12}$ & 8\\
R44 & \ce{C2H5 + H -> C2H4 + H2} & 2.9 & 300--2000 & $7.0 \times 10^{-11}$ & 6\\
R45& \ce{C2H5 + H -> CH3 + CH3} & 1.9 & 10--300 & $5.99 \times 10^{-11}$ & 8\\
R46 & \ce{CH3C2H + H + M -> CH3 + C2H2 + M} &2.3 &50--200& $8.89\times10^{-29} \left(\frac{T}{300}\right)^{-2.0} \exp\left(\frac{-1220}{T}\right)$ & 2\\
R47 & \ce{C4H + CH4 -> C4H2 + CH3} & 1.0 & 300--1000 & $2.98 \times 10^{-13} \left( \frac{T}{300} \right)^{2.72} \exp\left(\frac{-371}{T}\right)$ & 9\\
R48& \ce{C4H + CH4 -> C4H2 + CH3}& 1.0&50--200&$1.7\times10^{-12}\exp\left(\frac{-619}{T}\right)$ & 2\\
R49 &\ce{C4H + CH4 -> C4H2 + CH3}& 1.0&200--300&$1.6\times10^{-11}\exp\left(\frac{-610}{T}\right)$ & 10\\
R50 &\ce{C4H + CH4 -> C4H2 + CH3}& 1.0&300--1000&$2.98\times10^{-13}\left(\frac{T}{300}\right)^{2.72}\exp\left(\frac{-371}{T}\right)$ & 9 \\
R51 & \ce{C4H + C2H6 -> C4H2 + C2H5} & 1.1 & 50--200 & $2.87 \times 10^{-11} \left( \frac{T}{300} \right)^{-1.24} \exp\left(\frac{-25.6}{T}\right)$ & 2 \\
R52 & \ce{C4H2 + H + M -> C4H3 + M}&1.3&50--200&$3.30\times10^{-30}\exp\left(\frac{-740}{T}\right)$&2\\
R53&\ce{C4H3 + H -> C2H2 + C2H2}&5.5&50--200&$3.3\times10^{-12}$&2\\
R54&\ce{C4H3 + H -> H2 + C4H2}&3.2&10--800&$8.3\times10^{-11}$&1\\
R55&\ce{C4H3 + H -> C4H4}&4.8&50--200&$3.47\times10^{-27}\left(\frac{T}{300}\right)^{-1.3}$&\\
R56 & \ce{CN + CH4 -> HCN + CH3} & 0.9 & 10--160 & $5.73 \times 10^{-12} \exp\left(\frac{-675}{T}\right)$ & 2\\
R57 & \ce{CN + CH4 -> HCN + CH3} & 0.9 & 160--1500 & $6.2 \times 10^{-12} \exp\left(\frac{-735}{T}\right)$ & 6\\
R58 & \ce{CN + C2H2 -> H + HC3N} & 0.9 & 50--200 & $2.71 \times 10^{-10} \left( \frac{T}{300} \right)^{-0.52} \exp\left(\frac{-20}{T}\right)$ & 2 \\
R59 & \ce{CN + HCN -> C2N2 + H} & 0.5 & 50--200 & $4.3 \times 10^{-13} \left( \frac{T}{300} \right)^{1.71} \exp\left(\frac{-770}{T}\right)$ & 2\\

R60&\ce{HC2N2 + H -> HCN + HCN}& --&50-200& $1.7\times10^{-13}\exp \left(\frac{-110}{T}\right)$& 2\\
R61& \ce{H2C3N + H -> HCN + C2H2}&3.1&50--200&$1.5\times10^{-11}$& \\
R62&\ce{HC3N + H + M -> H2C3N + M}&1.4& 50--200 &$4.19\times 10^{-36}\exp\left(\frac{-740}{T}\right) n_\mathrm{tot}$&2\\
R63 & \ce{N(2D) + CH4 -> NH + CH3} & .. & 50--200 & $1.60 \times 10^{-10} \exp\left(\frac{-267}{T}\right)$ &11, 2\\
R64&\ce{N(2D) + NH3 -> NH + NH2} & --&200--300 & $5.00 \times 10^{-11}$ &7, 11
\\
R65&\ce{N + NH3 -> N2H + H2} & --&710--930 & $7.42 \times 10^{-12} \exp\left(-\frac{4830}{T}\right)$ & 7 \\
R66& \ce{NH + N -> N2 + H} & 6.4 & 10--500 & $5.00 \times 10^{-11} \left( \frac{T}{300} \right)^{0.1}$ \\
R67&\ce{NH + NH2 -> N2H2 + H} &1.14& 200--2500 & $6.98 \times 10^{-10} \left(\frac{T}{300}\right)^{-0.27} \exp\left(\frac{39}{T}\right)$ & 12 \\
R68&\ce{NH + NH2 -> NH3 + N} & 1.18 & -- & $1.59 \times 10^{-20} \left(\frac{T}{300}\right)^{2.46} \exp\left(-\frac{54}{T}\right)$ & 12 \\
R69&\ce{NH + N2H2 -> NH2 + N2H} & --&300--2000 & $4.00 \times 10^{-18} \left(\frac{T}{300}\right)^{2.0} \exp\left(\frac{600}{T}\right)$ & 7 \\
R70&\ce{NH + H2CN -> HCN + NH2} & --& -- & $3.0 \times 10^{-11}$ &7\\
R71&\ce{NH2 + CH4 -> CH3 + NH3} & 0.08& 300--2000 & $7.80 \times 10^{-12} \exp\left(-\frac{4680}{T}\right)$ & 7 \\
R72 & \ce{NH2 + C2H -> NH + C2H2}&0.76&--&$4.00\times10^{-11}$ &7\\
R73 & \ce{NH2 + C2H3 -> NH3 + C2H2} &3.1&--&$5.00\times10^{-11}$&7\\
R74 & \ce{NH2 + C2H4 -> NH3 + C2H3}&-0.18&1640--1940&$8.80\times10^{-12}\exp\left(\frac{-5170}{T}\right)$&7\\
R75&\ce{NH2 + C2H5 -> NH3 + C2H4} & 3.06 & -- & $3.15 \times 10^{-11}$ & 7 \\
R76&\ce{NH2 + C2H6 -> NH3 + C2H5} & 0.24&1500--2100 & $1.61 \times 10^{-11} \exp\left(-\frac{5770}{T}\right)$ & 7 \\
R77&\ce{NH2 + H2 -> NH3 + H} & 0.10&400--2200 & $2.49 \times 10^{-20} \left(\frac{T}{300}\right)^{2.6} \exp\left(-\frac{3007}{T}\right)$ & 7 \\
R78&\ce{NH2 + H2CN -> NH3 + HCN} & -- & -- & $2.29 \times 10^{-8} \left(\frac{T}{300}\right)^{-1.06} \exp\left(-\frac{60.8}{T}\right)$ & 13 \\
R79&\ce{NH2 + NH2 -> NH3 + NH} &0.53& 300--2500 & $9.35 \times 10^{-24} \left(\frac{T}{300}\right)^{3.53} \exp\left(-\frac{278}{T}\right)$ & 7 \\
R80&\ce{NH2 + NH2 -> N2H2 + H2} &1.57& 500--2250 & $2.89 \times 10^{-16} \left(\frac{T}{300}\right)^{1.02} \exp\left(-\frac{5930}{T}\right)$ & 7 \\
R81&\ce{NH2 + N2H2 -> NH3 + N2H} &-- &300--2000 & $3.0 \times 10^{-18} \left(\frac{T}{300}\right)^{1.94} \exp\left(\frac{580}{T}\right)$ & 7 \\
R82&\ce{NH2 + N2H3 -> NH3 + N2H2} &--& 600--2500 & $1.53 \times 10^{-18} \left(\frac{T}{300}\right)^{1.94} \exp\left(\frac{580}{T}\right)$ &7\\
R83&\ce{NH2 + NO -> N2H + OH} &-- &400--1900 & $7.28 \times 10^{-14} \left(\frac{T}{300}\right)^{0.29} \exp\left(\frac{438}{T}\right)$ & 7 \\
R84&\ce{NH2 + NO2 -> N2O + H2O} & 3.89&300--910 & $2.56 \times 10^{-8} \left(\frac{T}{300}\right)^{-1.44} \exp\left(-\frac{135}{T}\right)$ & 7 \\
R85&\ce{NH2 + O -> H2 + NO} & 3.52&298--3000 & $8.30 \times 10^{-12}$ & 7 \\
R86&\ce{NH2 + O3 -> H2O + NO2} & 5.51& 200--300 & $4.30 \times 10^{-12} \exp\left(-\frac{930}{T}\right)$ & 7 \\
R87&\ce{NH2 + OH -> H2O + NH} &1.06& 250--3000 & $1.49 \times 10^{-16} \left(\frac{T}{300}\right)^{1.5} \exp\left(\frac{230}{T}\right)$ & 7 \\
R88&\ce{NH3 + CH -> NH2 + CH2} & -0.26&300--1300 & $7.24 \times 10^{-11} \exp\left(\frac{317}{T}\right)$ & 7 \\
R89&\ce{NH3 + CN -> HCN + NH2} & 0.76&294--761 & $1.52 \times 10^{-11} \exp\left(\frac{180}{T}\right)$ & 7 \\
R90&\ce{NH3 + OH -> NH2 + H2O} & 0.53&300--2000 & $7.49 \times 10^{-14} \left(\frac{T}{300}\right)^{0.56} \exp\left(-\frac{730}{T}\right)$ & 14 \\
R91&\ce{N2H + NH2 -> N2 + NH3} &--& -- & $1.53 \times 10^{-18} \left(\frac{T}{300}\right)^{1.94} \exp\left(\frac{580}{T}\right)$ &7\\
R92&\ce{N2H2 + CH3 -> N2H + CH4} &-- &300--2000 & $2.66 \times 10^{-18} \left(\frac{T}{300}\right)^{1.87} \exp\left(-\frac{1495}{T}\right)$ & 7 \\
R93&\ce{N2H2 + H -> N2H + H2} &-- &300--2000 & $7.97 \times 10^{-16} \left(\frac{T}{300}\right)^{1.5} \exp\left(-\frac{795}{T}\right)$ & 7 \\

R94&\ce{N2H2 + O -> N2H + OH} &-- &300--2000 & $5.50 \times 10^{-16} \left(\frac{T}{300}\right)^{1.5} \exp\left(-\frac{250}{T}\right)$ & 7 \\
R95&\ce{N2H2 + OH -> N2H + H2O} &--& 300--2000 & $4.00 \times 10^{-18} \left(\frac{T}{300}\right)^{2.0} \exp\left(\frac{600}{T}\right)$ & 7 \\
R96&\ce{N2H3 + NH -> N2H2 + NH2} & -- & -- & $2.70 \times 10^{-12}$ &  \\
R97&\ce{NO + NH2 -> N2 + H2O} & 5.34&300--2000 & $7.90 \times 10^{-9} \left(\frac{T}{300}\right)^{-1.1} \exp\left(-\frac{98}{T}\right)$ & 7 \\

R98&\ce{H + NH2 + M -> NH3 + M} &4.62& 1--1000& $\displaystyle5.38\times10^{-25}\left(\tfrac{T}{300}\right)^{-2.0}$ &7, 15\\
R99 & \ce{H + HCN + M -> H2CN + M} & 1.1 & 50--200 & $3.3 \times 10^{-30} \left( \frac{T}{300} \right)^{-2.73} \exp\left(\frac{-3860}{T}\right)$ & 2\\
R100 & \ce{H + H2CN -> H2 + HCN} & 3.4 & 10--300 & $6.0 \times 10^{-11}$ & 8\\
R101 & \ce{H + C2N2 + M -> HCN2 + M}& --& 50--200& $1.0\times10^{-28}\exp \left(\frac{-740}{T}\right)$& 2\\
R102&\ce{H2 + N + M -> NH2 + M}&3.01& 1--1000& $1.0\times10^{-36}$ & 7\\
R103&\ce{HO2 + NH2 -> NH3 + O2} & 2.38&300--2000 & $2.75 \times 10^{-20} \left(\frac{T}{300}\right)^{1.55} \exp\left(-\frac{1020}{T}\right)$ & 7, 16\\

R104&\ce{O + NH3 -> OH + NH2} & -0.17&500--2500 & $1.60 \times 10^{-11} \exp\left(-\frac{3670}{T}\right)$ & 7 \\

R105& \ce{OH + CH3 -> ^1CH2 + H2O} & -- & 150--1200 & $1.0 \times 10^{-10} \left( \frac{T}{300} \right)^{-0.91} \exp\left(\frac{-275}{T}\right)$ &7 \\
R106& \ce{OH + CH3 -> CH3O + H} & -- & 298--2000 & $2.0\times10^{-14}\exp\left( \frac{-6990}{T}\right)$ & 6 \\
R107& \ce{OH + CH3 -> CH2OH + H} &--& 298--2000 & $1.2 \times 10^{-12} \exp\left(\frac{-2760}{T}\right)$ & 6\\
R108 & \ce{OH + CH3 -> H2 + H2CO} & -- & 298--2000 & $5.3\times 10^{-15} \exp\left(\frac{-2760}{T}\right)$ &6\\
R109& \ce{OH + NH -> O + NH2} & -0.36 & 298--3000& $1.66\times10^{-12} \left(\frac{T}{300}\right)^{0.1}\exp\left(\frac{-5800}{T}\right)$&7\\
R110& \ce{OH + NH -> O + NH2} & -0.36 & 298--3000& $1.66\times10^{-12} \left(\frac{T}{300}\right)^{0.1}\exp\left(\frac{-5800}{T}\right)$&7\\
R111&\ce{OH + HCN -> NH2 + CO} &0.99& 500--2500 & $1.07 \times 10^{-13} \exp\left(-\frac{5890}{T}\right)$ &7 \\

R112&\ce{H2O + CH3 -> CH4 + OH} &-- &50--200 & $1.22\times10^{-14} \left(\frac{T}{300}\right)^{2.9} \exp\left(\frac{-7480}{T}\right)$ &\\
R113&\ce{HCO + CH3 -> CO + CH4} &-- &50--200 & $2.01\times10^{-10}$ &\\
R114&\ce{H2CO + CH3 -> CH4 + HCO} &-- &50--200 & $6.79\times10^{-12} \exp\left(\frac{-4450}{T}\right)$ &\\
R115&\ce{H2CO + CH3 -> CH4 + HCO} &-- &300--2000 & $1.12\times10^{-14} \left(\frac{T}{300}\right)^{3.36} \exp\left(\frac{-2170}{T}\right)$ &\\
R116&\ce{CH3O + CH3 -> H2CO + CH4} &--& 50--200 & $4.00\times10^{-11}$ & \\
R117&\ce{CH2OH + CH3 -> H2CO + CH4} &-- &50--200 & $4.00\times10^{-12}$ &\\

\hline
\\[0.01mm]

\multicolumn{6}{p{\dimexpr\linewidth-2\tabcolsep}}{%
  \footnotesize
  \justifying
  \noindent
  {\bf References}: Reaction network based on the chemical scheme of \citet{yung_photochemistry_1984}. The temperature ranges and rate constants are taken from KIDA \citep{wakelam_kinetic_2012}. The formation enthalpy values used to calculate the reaction energies are taken from the Argonne Active Thermochemical Tables \citep{ruscic_active_2022} and NIST \citep{linstrom_nist_2001}. For reactions involving a third body the Modified Arrenhius equation for the low pressure condition is multiplied by the total number density of the gas. If any data is taken from other resources they are listed in the reference column, $^1$ \citealt{harada_modeling_2008}, $^2$ \citealt{hebrard_how_2009}, $^3$ \citealt{douglas_low_2018} , $^4$ \citealt{gannon_state_2010} , $^5$  \citealt{yung_photochemistry_1984}, $^6$ \citealt{baulch_evaluated_2005},$^7$ \citealt{tsai_comparative_2021}, $^8$ \citealt{loison_interstellar_2014}, $^9$ \citealt{huo_direct_2011}, $^{10}$ \citealt{berteloite_low_2010}, $^{11}$ \citealt{herron_evaluated_1999},$^{12}$ \citealt{klippenstein_thermal_2009},$^{13}$ \citealt{yelle_formation_2010}, $^{14}$ \citealt{samu_determination_2018}, $^{15}$ \citealt{pagsberg_pulse_1979}, $^{16}$ \citealt{sumathi_quantum_1996},
   }
\end{longtable}


\bsp	
\label{lastpage}
\end{document}